%% file: DTD.tex
\documentclass[twocolumn,appendixfloats,tighten]{aastex6}
\usepackage{newtxtext,newtxmath}
\usepackage[T1]{fontenc}
\usepackage{ae,aecompl}
\usepackage{amsmath}	% Advanced maths commands
\usepackage{amssymb}	% Extra maths symbols

\usepackage{ctable}
\usepackage{url}
\usepackage{xspace}
\usepackage[normalem]{ulem}
\usepackage{hyperref}
\usepackage[all]{hypcap}
\usepackage{graphicx}
\usepackage{color}

\def\msun{{\rm\,M_\odot}}

\def\msun{{\rm\,M_\odot}}

\newcommand{\be}{\begin{equation}}
\newcommand{\ee}{\end{equation}}

\newcommand{\tmin}{t_{\rm min}}
\def\h2{${\rm\,H_2}$}

\usepackage{color}

\begin{document}

\title{Measuring the delay time distribution of binary neutron stars. II. Using the redshift distribution from third-generation gravitational wave detectors network}
%Measuring the delay time distribution of binary neutron stars with third-generation gravitational-wave detectors

\author{Mohammadtaher Safarzadeh\altaffilmark{1,2}, Edo Berger\altaffilmark{1}, Ken K.~Y.~Ng\altaffilmark{3,4}, Hsin-Yu Chen\altaffilmark{5},  Salvatore Vitale\altaffilmark{3,4}, Chris Whittle\altaffilmark{3,4}, Evan Scannapieco\altaffilmark{2}}

\altaffiltext{1}{Center for Astrophysics | Harvard \& Smithsonian, 60 Garden Street, Cambridge MA 02138, USA \href{mailto:msafarzadeh@cfa.harvard.edu}{msafarzadeh@cfa.harvard.edu}}
\altaffiltext{2}{School of Earth and Space Exploration, Arizona State University Tempe AZ 85287, USA}
\altaffiltext{3}{LIGO, Massachusetts Institute of Technology, 185 Albany Street, Cambridge MA 02139, USA}
\altaffiltext{4}{Department of Physics and Kavli Institute for Astrophysics and Space Research, MIT, 77 Massachusetts Avenue, Cambridge MA 02139, USA}
\altaffiltext{5}{Black Hole Initiative, Harvard University, 20 Garden Street, Cambridge MA 02138, USA}

\begin{abstract}
We investigate the ability of current and third-generation gravitational wave (GW) detectors to determine the delay time distribution (DTD) of binary neutron stars (BNS) through a direct measurement of the BNS merger rate as a function of redshift.  We assume that the DTD follows a power law distribution with a slope $\Gamma$ and a minimum merger time $t_{\rm min}$, and also allow the overall BNS formation efficiency per unit stellar mass to vary. By convolving the DTD and mass efficiency with the cosmic star formation history, and then with the GW detector capabilities, we explore two relevant regimes.  First, for the current generation of GW detectors, which are only sensitive to the local universe, but can lead to precise redshift determinations via the identification of electromagnetic counterparts and host galaxies, we show that the DTD parameters are strongly degenerate with the unknown mass efficiency and therefore cannot be determined uniquely.  Second, for third-generation detectors such as Einstein Telescope (ET) and Cosmic Explorer (CE), which will detect BNS mergers at cosmological distances, but with a redshift uncertainty inherent to GW-only detections ($\delta(z)/z\approx 0.1z$), we show that the DTD and mass efficiency can be well-constrained to better than 10\% with a year of observations.  This long-term approach to determining the DTD through a direct mapping of the BNS merger redshift distribution will be supplemented by more near term studies of the DTD through the properties of BNS merger host galaxies at $z\approx 0$ \citep{SB19}.
\end{abstract}

\section{Introduction}

The joint gravitational wave (GW) and electromagnetic (EM) detections of the binary neutron star (BNS) merger, GW170817 \citep{Abbott:2017it}, marked the dawn of multi-messenger astronomy. As the current generation of GW detectors increases in sensitivity and number, the local rate of BNS mergers will soon be determined accurately for the first time, providing initial insight into the formation channels of these binaries.  Currently, the BNS merger rate is weakly constrained by the single detection of GW170817 ($1540^{+3200}_{-1220}$ Gpc$^{-3}$ yr$^{-1}$; \citealt{Collaboration:2017kt}), by the small known sample of Galactic BNS systems ($21^{+28}_{-14}$ Myr$^{-1}$; \citealt{Kim:2015bi}), and by the beaming-corrected rate of short gamma-ray bursts ($270^{+1580}_{-180}$ Gpc$^{-3}$ yr$^{-1}$; \citealt{fbm+15}).  These rates are in broad agreement, but the uncertainties from all methods span at least two orders of magnitude.

Still, even when the local BNS merger rate is well determined, the more fundamental distribution of merger delay times may not be.  The delay time distribution (DTD) encodes the time span between the formation of the BNS system (or alternatively the time since the formation of the parent stars) until the two neutron stars merge through the emission of gravitational waves.  The DTD therefore provides fundamental insight into the evolutionary processes that govern the initial separation of the binaries, including poorly-understood effects such as common envelope evolution.

The DTD is usually parametrized as a power law distribution above some minimum merger timescale, $\tmin$, based on the following arguments: after the BNS formation, the binary's orbit decays due to the emission of gravitational waves on a timescale that depends on the initial semi-major axis ($a$) as $t\propto a^4$. Therefore the resulting distribution of the merger times depends on the distribution of initial semi-major axes, $dN/da\propto a^{-\beta}$. The initial semi-major axis distribution of the O/B stellar progenitors is assumed to follow a power law $dN/da\propto a^{-1}$. If the binary experiences a common envelope phase, then the distribution becomes steeper. Therefore, the expected merger times follow $dN/dt_{\rm merge}\propto t^{-\beta/4-3/4}$, where we define $\Gamma\equiv -\beta/4-3/4$ \citep{Belczynski:2018vr}.

Insight on the form of the DTD has been gained from studies of the small population of Galactic BNS systems \citep{VignaGomez:2018th}, from the properties of SGRB host galaxies \citep{Zheng:2007hl,OShaughnessy:2008bm,lb10,fbc+13,Behroozi:2014bp,Berger14}, and from arguments related to $r$-process enrichment \citep{Matteucci:2014jta,Komiya:2014ie,vandeVoort:2015jw,Shen:2015gc,Cote:2018gj,Hotokezaka2018IJMPD,Safarzadeh:2018ub,Safarzadeh:2019dd}. 
These results point to the need for a fast merging channel if BNSs are assumed to be the primary source of $r$-process enrichment in the universe, which suggest that $\tmin$ may be rather small, $\lesssim 0.1$ Gyr or the slope of the DTD is steep. Population synthesis models have also made various predictions for the values of $\Gamma$ and $\tmin$, but those are dependent on uncertain binary evolution processes \citep{Dominik:2012cwa}.  We stress that the local BNS merger rate in itself cannot fully characterize the DTD since it also depends on an additional unknown parameter, the efficiency of BNS formation per unit stellar mass, $\lambda$. 

In a recent paper, \citet{SB19} (hereafter, Paper I) showed that the mass distribution of BNS merger host galaxies at $z\approx 0$ can provide insight on $\Gamma$ and $\tmin$ with a sample size of $\mathcal{O}(10^2-10^3)$. This is based on the fact that, on average, galaxy star formation histories (SFH) depend on their mass, and hence the convolution of the DTD and SFH leads to a specific prediction about the mass function of BNS merger host galaxies. Such an observational approach to determining the DTD is only feasible in the local universe due to the required detection of EM counterparts that will in turn lead to the identification of the host galaxies.  It is anticipated that Advanced LIGO/Virgo, joined by KAGRA\footnote{https://gwcenter.icrr.u-tokyo.ac.jp/en} and IndIGO\footnote{http://www.gw-indigo.org/tiki-index.php}, can produce the required sample size within the next two decades. 

Here, we instead explore how the DTD can be determined by directly observing the redshift distribution of BNS mergers well beyond the local universe.  Mapping the rate of BNS mergers as a function of redshift can break the degeneracy between the shape of the DTD ($\Gamma$ and $\tmin$) and the BNS mass efficiency ($\lambda$) when comparing to the cosmic star formation history.  This approach requires an order of magnitude increase in GW detector sensitivity to detect BNS mergers at cosmological distances.  Such an improvement is expected for third-generation ground-based observatories such as Einstein Telescope\footnote{http://www.et-gw.eu} (ET;   \citealt{Punturo:2010jf}) and Cosmic Explorer\footnote{http://www.cosmicexplorer.org} (CE; \citealt{Abbott:2017ie}).  However, at these distances, it is unlikely that EM counterparts will be detected for the majority of events, and therefore the distance (redshift) information will rely directly on the GW signal itself.  We explore how the inherent distance-inclination degeneracy affects the ability to determine the DTD.

The structure of the paper is as follows: In \S\ref{sec:method} we delineate the method of estimating the observed redshift distribution of BNS mergers as a function of DTD and mass efficiency, for different GW interferometer networks; in \S\ref{sec:ligo} we show the results of DTD determination for existing GW detectors, which are only sensitive to the local universe; in \S\ref{sec:3g} we expand our analysis to a future network of ET and CE, including a determination of the expected redshift uncertainties, and show the resulting constraints on the DTD and mass efficiency.  We discuss some caveats and summarize the key results in \S\ref{sec:summary}.  We adopt the Planck 2015 cosmological parameters \citep{Collaboration:2016bk} where $\Omega_M=0.308$, $\Omega_\Lambda=0.692$, $\Omega_b=0.048$ are total matter, vacuum, and baryonic densities, in units of
the critical density, $\rho_c$, $H_0=67.8$ km s$^{-1}$ Mpc$^{-1}$ is the Hubble constant, and $\sigma_8=0.82$ is the variance of linear fluctuations on the 8 $h^{-1}$ Mpc scale.

\section{method}
\label{sec:method}

The BNS merger rate as a function of redshift is a convolution of the DTD with the cosmic star formation rate density:
\begin{align}
\dot{n}(z)=&\int_{z_b=10}^{z_b=z} \lambda\frac{dP_m}{dt}(t-t_b-t_{\rm min})\psi(z_b)\frac{dt}{dz}(z_b)dz_b,
\end{align}
where $dt/dz = -[(1+z) E(z) H_0]^{-1}$, and 
$E(z)=\sqrt{{\Omega}_{m,0}(1+z)^3+{\Omega}_{k,0}(1+z)^2+{\Omega}_{\Lambda}(z)}$.
Here, $\lambda$ is the currently unknown BNS mass efficiency (assumed not to evolve\footnote{Although the DTD for binary black holes is likely highly dependent on the metallicity, the DTD for BNS systems has been argued to be at most weakly dependent on  metallicity \citep{Dominik:2012cwa}.} with redshift) used as a free parameter that we try to recover alongside the parameters governing the DTD; $t_b$ is the time corresponding to the redshift $z_b$; $dP_m/dt$ is the DTD, parametrized to follow a power law distribution ($\propto t^{\Gamma}$) with a minimum delay time, $t_{\rm min}$ that refers to the time since birth of the ZAMS stars and not when the BNS system formed. 
Therefore, $t_{\rm min}$ corresponds to the sum of the nuclear lifetime of the lowest mass component of the binary system and the minimal gravitational delay that is induced by the existence of a minimal separation between the two newly born neutron stars. We also impose a maximum delay time of 10 Gyr for our fiducial case, although this does not affect our results, although we note that more than half of the observed BNS systems in the MW half merger times more than 10 Gyr \citep[][]{Pol2019A}. We adopt the cosmic star formation rate density\footnote{We neglect the uncertainties in the cosmic SFRD since the GW source redshift uncertainty (see Appendix~\ref{appendixB}) dominates the overall error budget at cosmological redshifts.} from \citet{Madau:2014gtb}:
\be
\psi(z)=0.015 \frac{(1+z)^{2.7}}{1+[(1+z)/2.9]^{5.6}}\,\, \msun\, {\rm yr^{-1}\, Mpc^{-3}}.
\ee 

\begin{figure}[t]
\hspace{-0.2in}
\centering
\includegraphics[width=1.05\columnwidth]{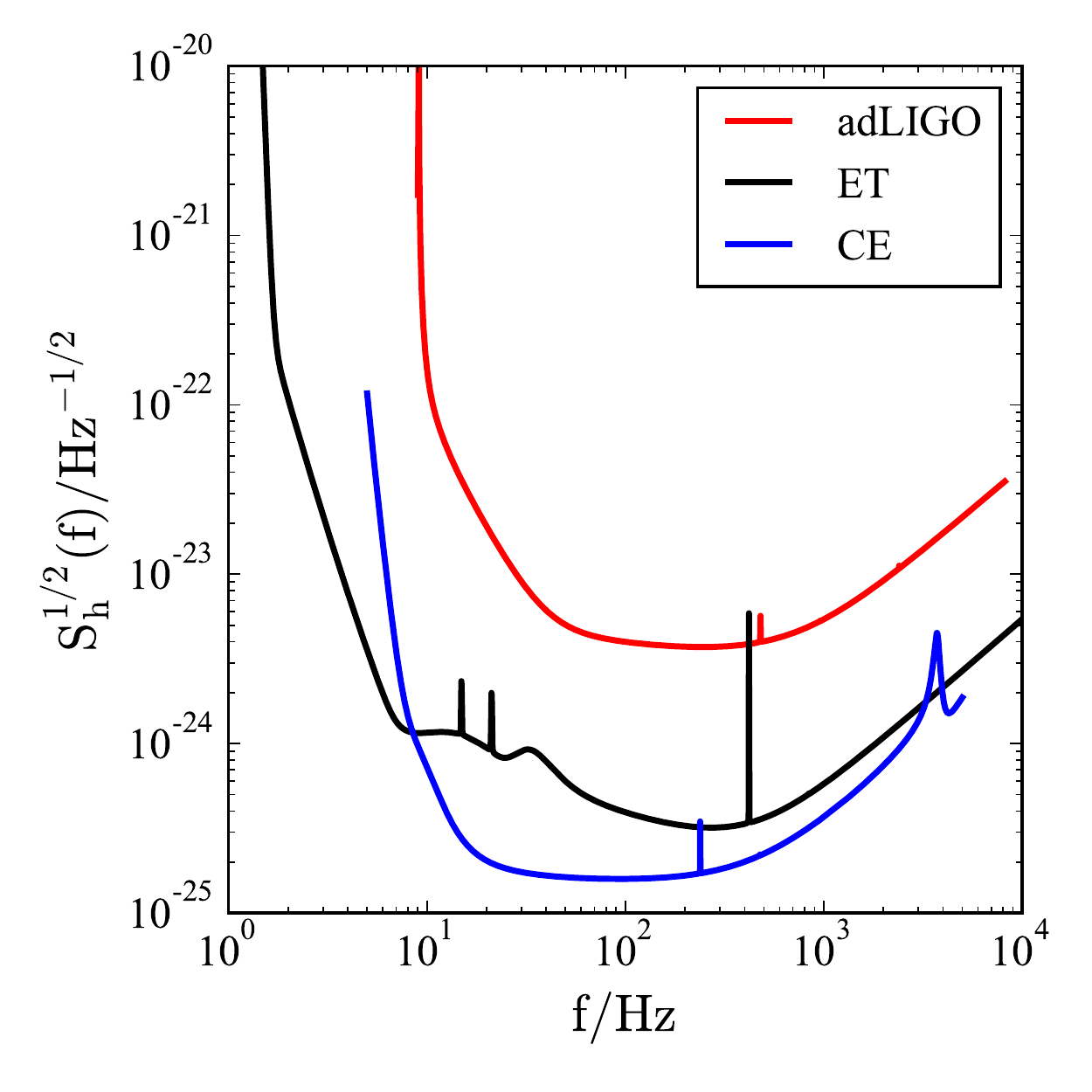}
\caption{Comparison of the noise curves of different GW interferometers studied in this work. Red, black, and blue lines correspond to Advanced LIGO, Einstein Telescope (ET), and Cosmic Explorer (CE), respectively.}
\label{f:gw}
\end{figure}

To determine the {\it observed} BNS merger rate as a function of redshift we need to consider the matched filtering signal-to-noise ratio as a function of GW detector sensitivity \citep{Finn96}:
\begin{equation} 
\label{eq:snr}
{\rho(z)}=8{\Theta}{\frac{r_{0}}{D_{L}}}{\left({\frac{{\mathcal{M}}_{z}}{1.2M_{\odot}}}\right)}^{5/6}\sqrt{\zeta(f_{\rm{max}})},
\end{equation}
where $\mathcal{M}_{z} = (1+z)\mathcal{M}$ is the redshifted chirp mass, $D_L$ is the luminosity distance, $\Theta$ is the orientation function, and
\begin{align} \label{eq:distance_r0_table}
{r_{0}^{2}}&\equiv{\frac{5}{192\pi}}\left({\frac{3G}{20}}\right)^{5/3}x_{7/3}\frac{M_{\odot}^{2}}{c^3},\nonumber\\
x_{7/3}&\equiv{\int_{0}^{\infty}\frac{df({\pi}M_{\odot})^{2}}{{({\pi}fM_{\odot})^{7/3}}S_{h}(f)}},\nonumber\\
{\zeta(f_{\rm{max}})}&\equiv{\frac{1}{x_{7/3}}}{\int_{0}^{2f_{\rm{max}}}\frac{df({\pi}M_{\odot})^{2}}{{({\pi}fM_{\odot})^{7/3}}S_{h}(f)}},
\end{align}
where $2f_{\rm{max}}$ is the wave frequency at which the inspiral detection template ends, $r_0$ denotes the characteristic distance sensitivity, and $S_{h}(f)$ is the detector's noise power spectral density. The intrinsic chirp mass, $\mathcal{M}$, is given in terms of the component masses by:
\begin{equation}
\mathcal{M}={\left(\frac{m_{1}m_{2}}{{(m_{1}+m_{2})}^{2}}\right)}^{3/5}(m_{1}+m_{2}).
\end{equation}
Here we assume that both neutron stars have mass of $m_1=m_2=1.4$  $\msun$.
The frequency at the end of the inspiral (taken to correspond to the innermost stable circular orbit) is:
\begin{equation}\label{eq:ISCO}
f_{\rm {max}}=\frac{785\text{ Hz}}{1+z}\left({\frac{2.8M_{\odot}}{M}}\right),
\end{equation}
where $M$ is the total mass of the binary. In Figure~\ref{f:gw} we show the sensitivity curves for Advanced LIGO, ET, and CE. The substantial reduction in noise amplitude for the third-generation detectors with respect to Advanced LIGO leads to an increase in the typical values of $r_0$ from $\approx 0.1$ to $\approx 1.5$ Gpc. Finally, the {\it observed} BNS merger rate as a function of redshift is given by:
\begin{equation} \label{eq:detectionrate}
R_D(z)= \frac{dV_c}{dz}\frac{\dot{n}(z)}{1+z} P_{\text{det}}(z),
\end{equation}
where $P_{\text{det}}(z)$ is defined in Appendix~\ref{CofTheta}, and the redshift derivative of the comoving volume is given by
$dV_c/dz=(4\pi c/H_0)[D_L^2/(1+z)^2E(z)]$.

\begin{figure}[t]
\hspace{-0.2in}
\centering
\includegraphics[width=1.05\columnwidth]{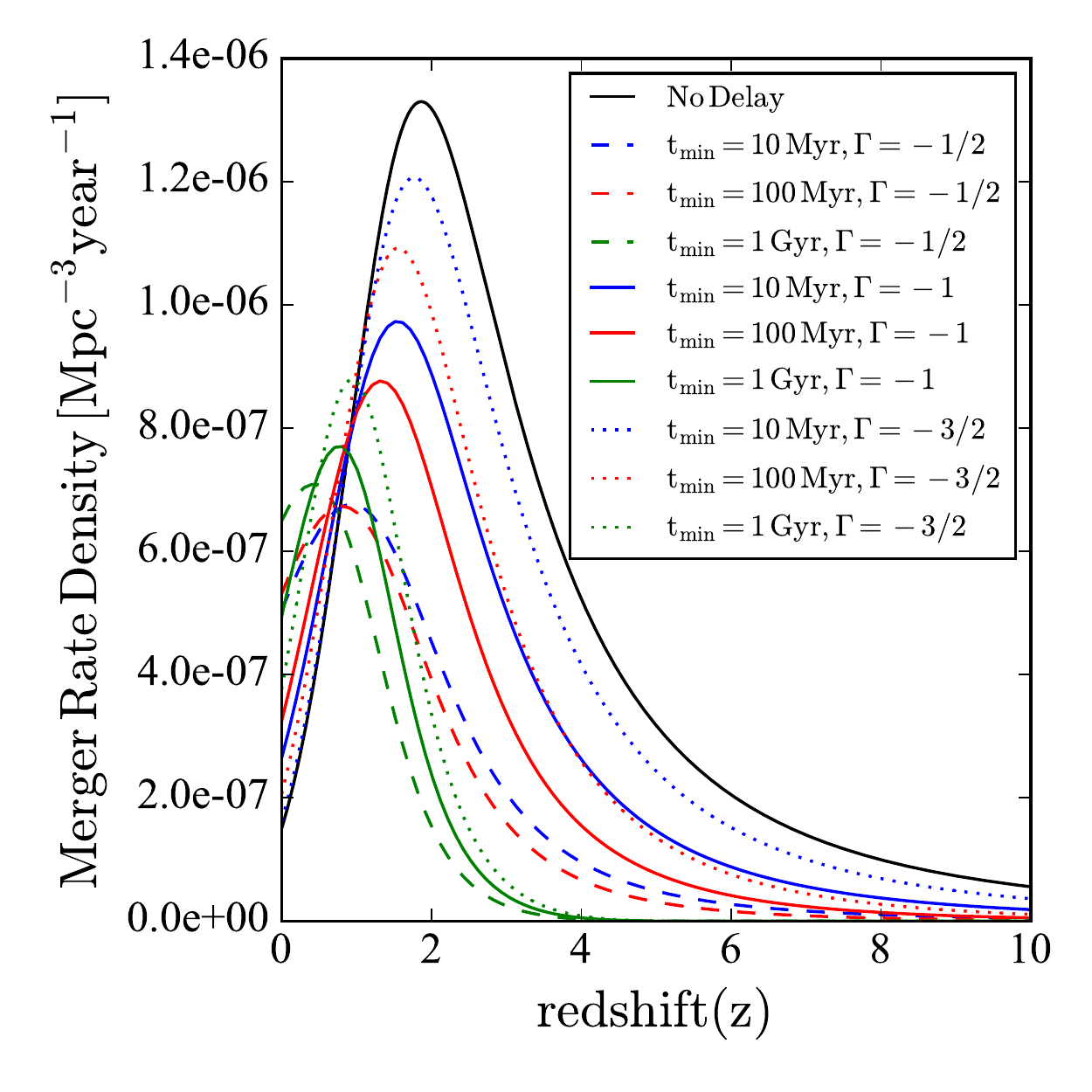}
\caption{The {\it intrinsic} redshift distribution of BNS mergers formed according to the cosmic star formation rate density, and with different DTDs spanning a range of $\Gamma$ and $\tmin$.  We assume a BNS mass efficiency of $\lambda=10^{-5}$ M$_{\odot}^{-1}$. For comparison, the solid black line shows the merger rate density in the absence of a delay.}
\label{f:all_models}
\end{figure}

In Figure~\ref{f:all_models} we show the {\it intrinsic} merger rate density, $\dot{n}(z)$, for nine different choices of the DTD, with $\Gamma=[-1.5,-1,-0.5]$ and $\tmin=[10,100,1000]$ Myr, and a fixed mass efficiency value of $\lambda=10^{-5}$ M$_\odot^{-1}$. For comparison, we also show the curve corresponding to no delay (i.e., the cosmic star formation rate density).  Clearly, DTDs that prefer longer delays result in a merger distribution that is skewed to lower redshifts, with a higher merger rate at $z\approx 0$, but with some degeneracy between $\Gamma$ and $\tmin$. However, since the value of $\lambda$ is not presently known, all of the DTDs can reproduce the same local rate by simply scaling $\lambda$ appropriately.  This is essentially why a local measurement of the merger rate cannot by itself constrain the DTD.

To explore how well current and third-generation GW detectors can determine the DTD, we inject a specific DTD model ($\Gamma$, $\tmin$, $\lambda$), generate the resulting redshift distribution with associated uncertainties, and then fit this distribution using an interpolation table that is based on the nine input DTDs. We fit for the input parameters using Markov Chain Monte Carlo (MCMC) sampling with {\em emcee}, a python based affine invariant sampler \citep{ForemanMackey:2013io}. The likelihood function is
$ln (L)=-\chi^2/2$, with $\chi^2=\L\sum_{i=0}^{i=N} (R_{D,i} - \hat{R_{D,i}})/\sigma_{t,i}^2$.  Here the summation is over all of the redshift bins; $R_{D,i}$ and $\hat{R_{D,i}}$ are the constructed and simulated detection rates at redshift bin $i$, respectively; $\sigma_{t,i}$ is the total error on the detection rate at redshift bin $i$, which is a combination of the Poisson error and the error due to the distance-inclination degeneracy from GW data, $\sigma_t^2=\sigma_p^2 +\sigma_z^2$; $\sigma_p=\sqrt{N}$, where $N$ is the expected number of events at a given redshift during the integrated observation time of length $T_{\rm obs}$; and the redshift uncertainty ($\sigma_z$) for each redshift bin is estimated based on the vertical distance from the mean expected detection rate to the upper envelope corresponding to when the detections' redshift are all biased high. We model the redshift uncertainty as $\delta z/z=0.1 z$ based on re-scaled simulations of binary black hole redshift uncertainty estimates as detailed in Appendix \ref{appendixB}. We adopt a flat prior distribution for all of our parameters in the $\log\lambda \in [-7,-3]$, $\log t_{\rm min} \in [1,3]$, and $\Gamma \in [-1.5,-0.5]$.

\section{results for Current GW Detectors}
\label{sec:ligo}

\begin{figure}
\hspace{-0.2in}
\centering
\includegraphics[width=1.05\columnwidth]{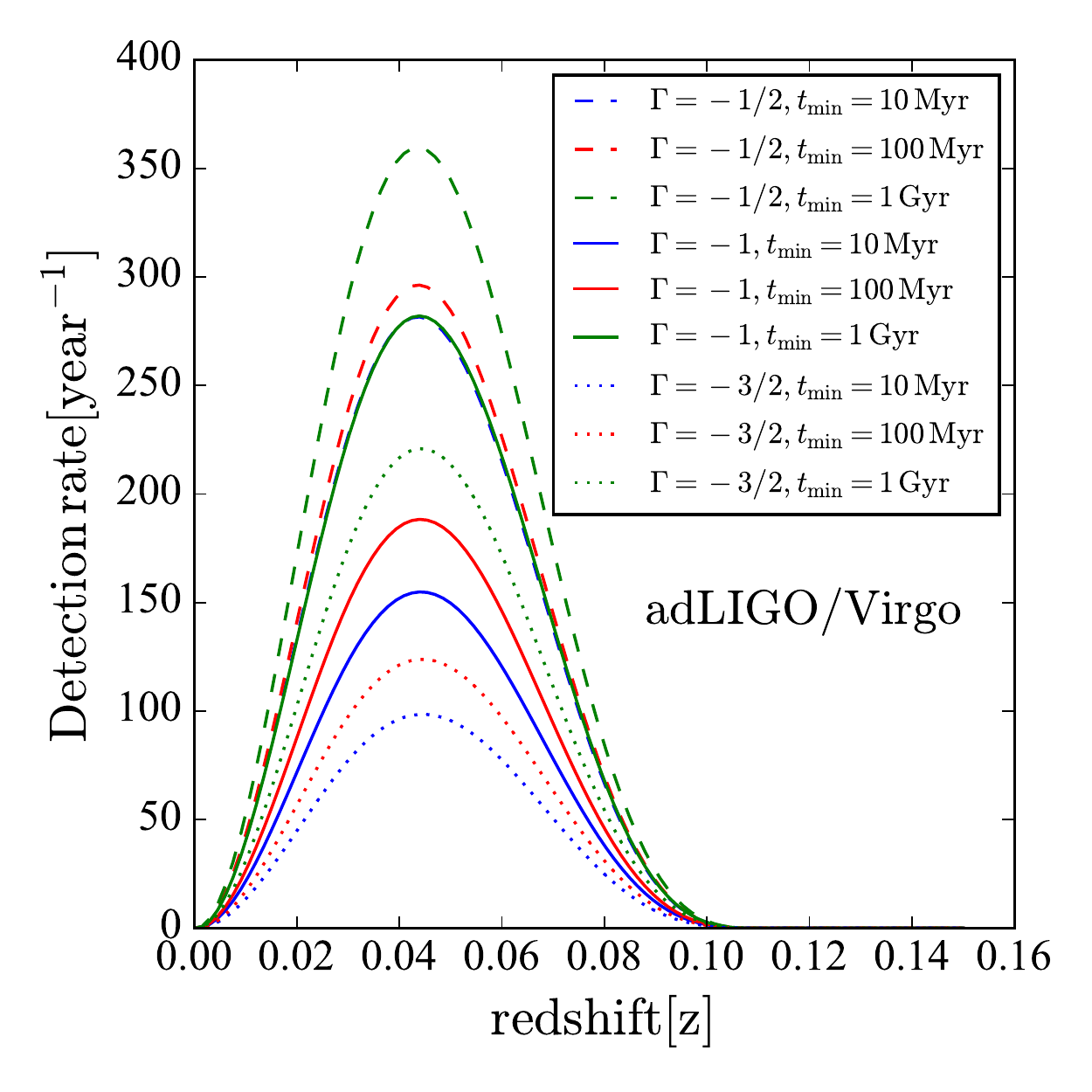}
\caption{ The expected detection rate as a function of redshift for Advanced LIGO, for the nine DTDs shown in Figure~\ref{f:all_models}. The detection PDFs are basically identical (modulo a scaling with the unknown value of $\lambda$) because Advanced LIGO can only detect BNS mergers in the local universe.  We consider a minimum signal-to-noise ratio of 8 for detection.}
\label{f:DRs}
\end{figure}

\begin{figure}
\hspace{-0.15in}
%\vspace{1cm}
\centering
\includegraphics[width=1.05\columnwidth]{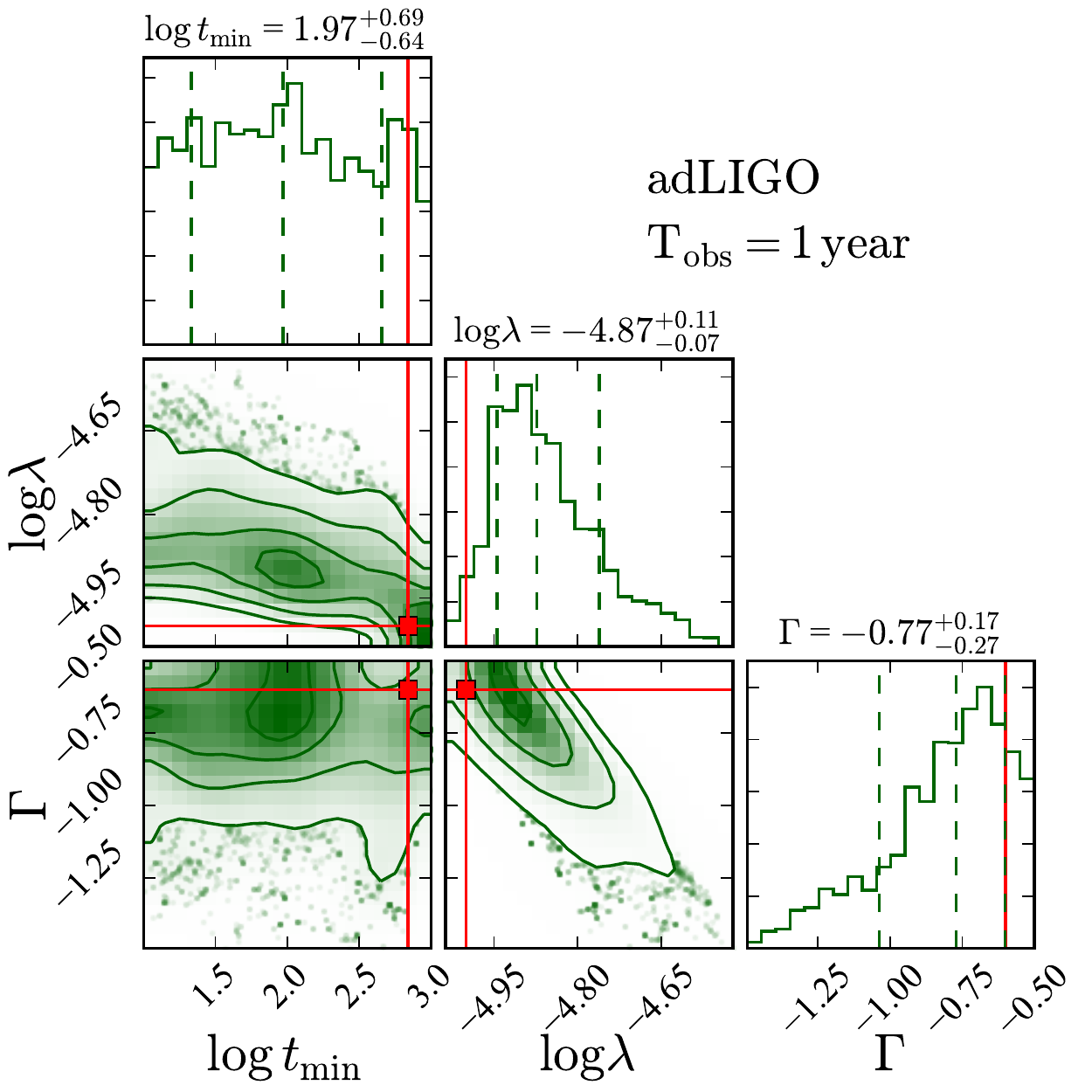}
\caption{Results of MCMC parameter estimation for a year of Advanced LIGO/Virgo operations at design sensitivity.  The red vertical lines and circles mark the input DTD model, while the green curves and contours show the posteriors of the model parameters.  The black lines show the median and range of 16th to 84th percentiles.  Here we assume that the redshifts are known precisely thanks to EM counterparts and host galaxy identifications.}
\label{f:mcmc_adLIGO}
\end{figure}

\begin{figure*}
%\setlength{\tabcolsep}{1mm}
%\hspace{0cm}
\vspace{1cm}
\begin{center}
\includegraphics[width=0.3\textwidth]{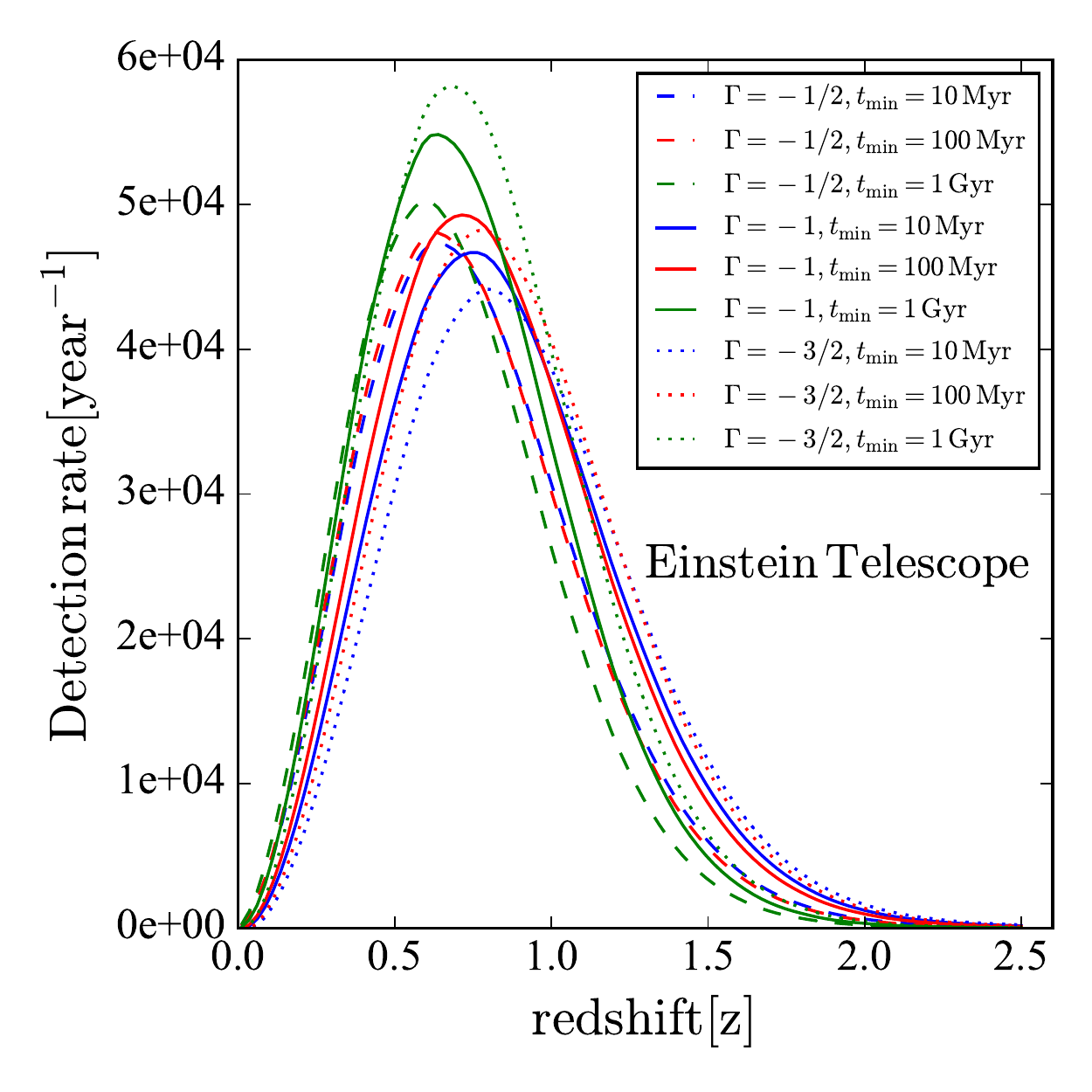}
\includegraphics[width=0.3\textwidth]{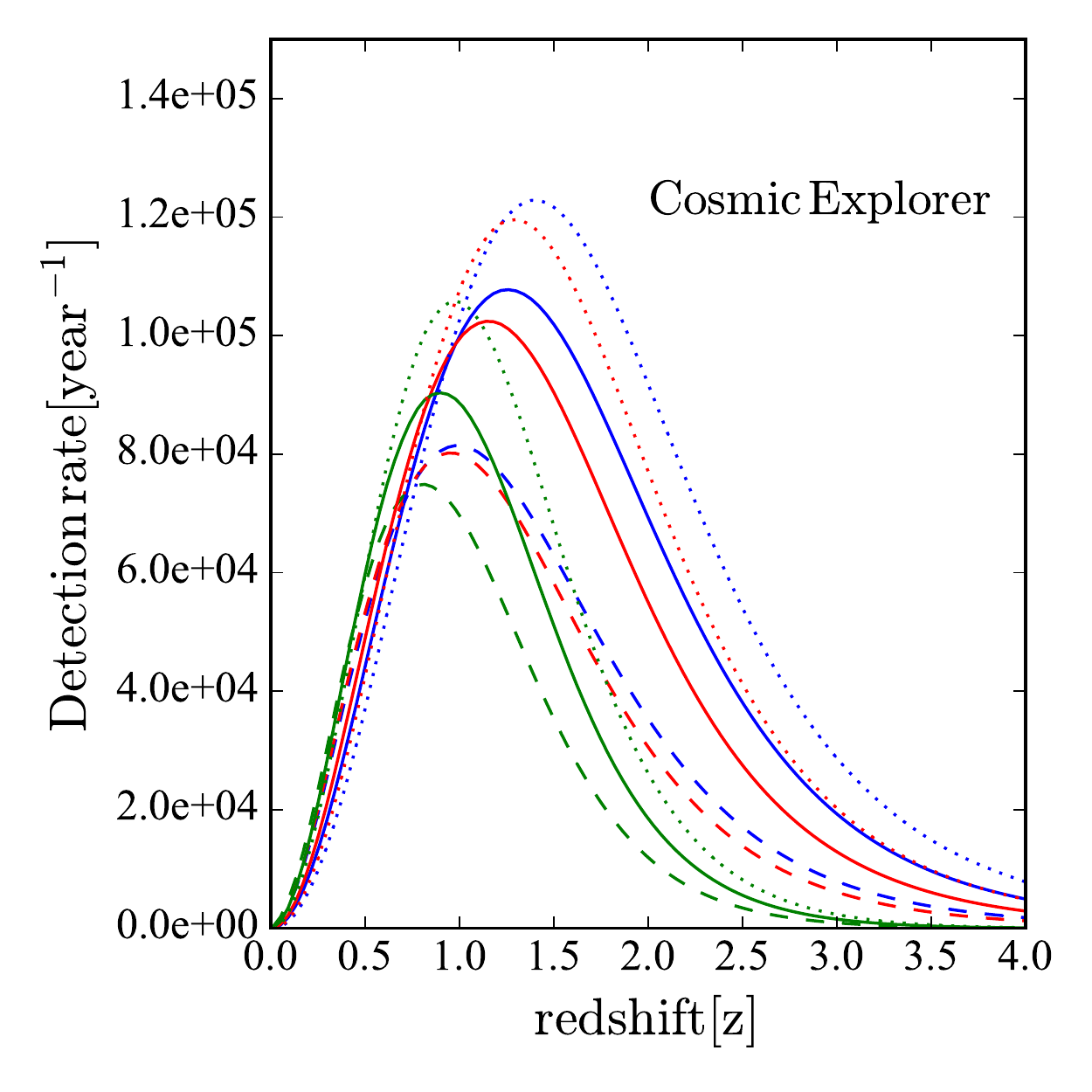}
\includegraphics[width=0.3\textwidth]{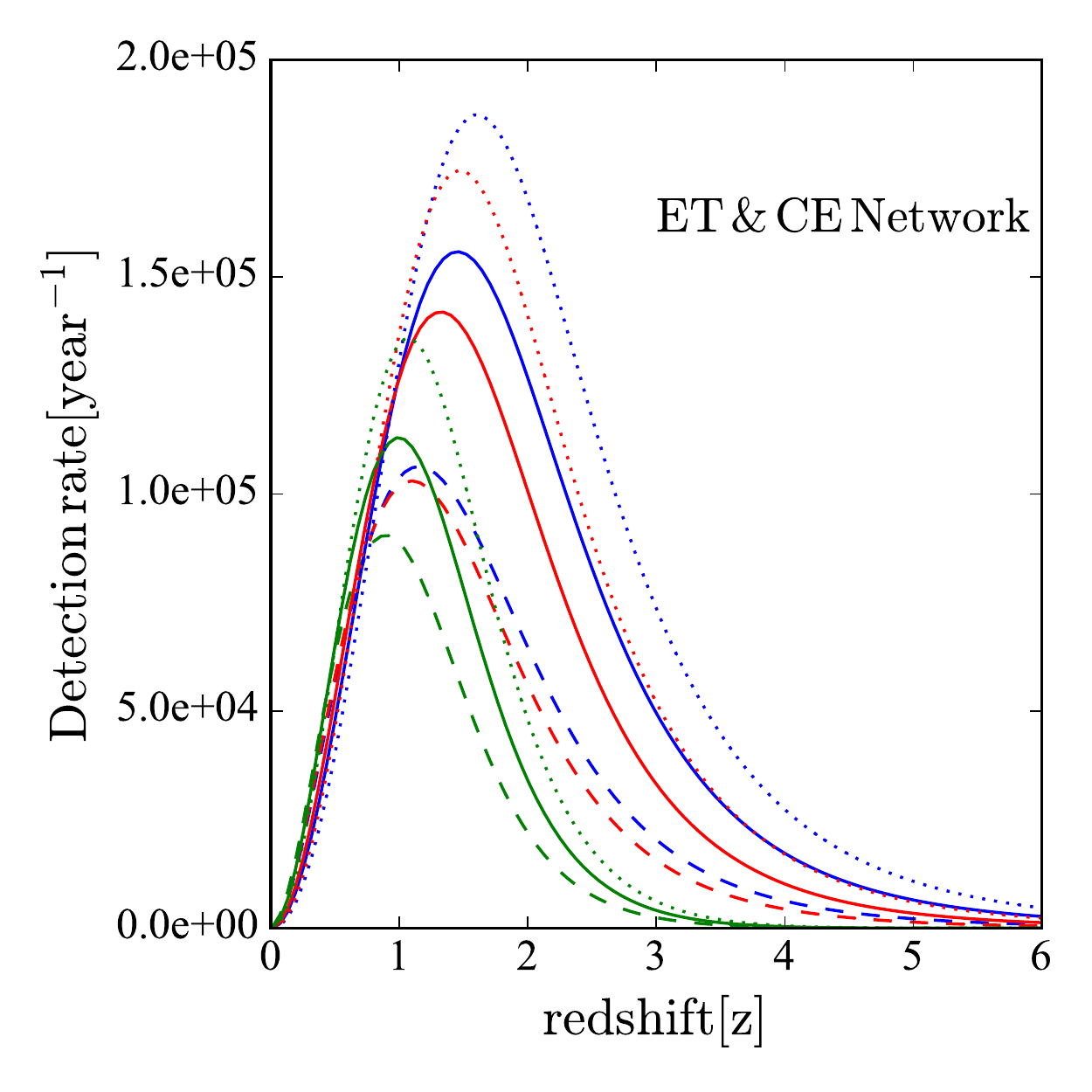}
\end{center}
\caption{The expected detection rate as a function of redshift for ET ({\it left}), CE ({\it middle}), and a network of ET+CE ({\it right}), for the nine DTDs shown in Figure~\ref{f:all_models}. Due to the ability of these third-generation detectors to detect BNS mergers at cosmological distances, the resulting redshift distributions are no longer fully degenerate.  The network of ET+CE not only leads to greater sensitivity, but also provides improved redshift determination compared to ET or CE alone (Appendix \ref{appendixB}). We consider a minimum signal-to-noise ratio of 8 for detection.}
\label{f:third_gen}
\end{figure*}

\begin{figure*}
\centering
\includegraphics[width=8cm]{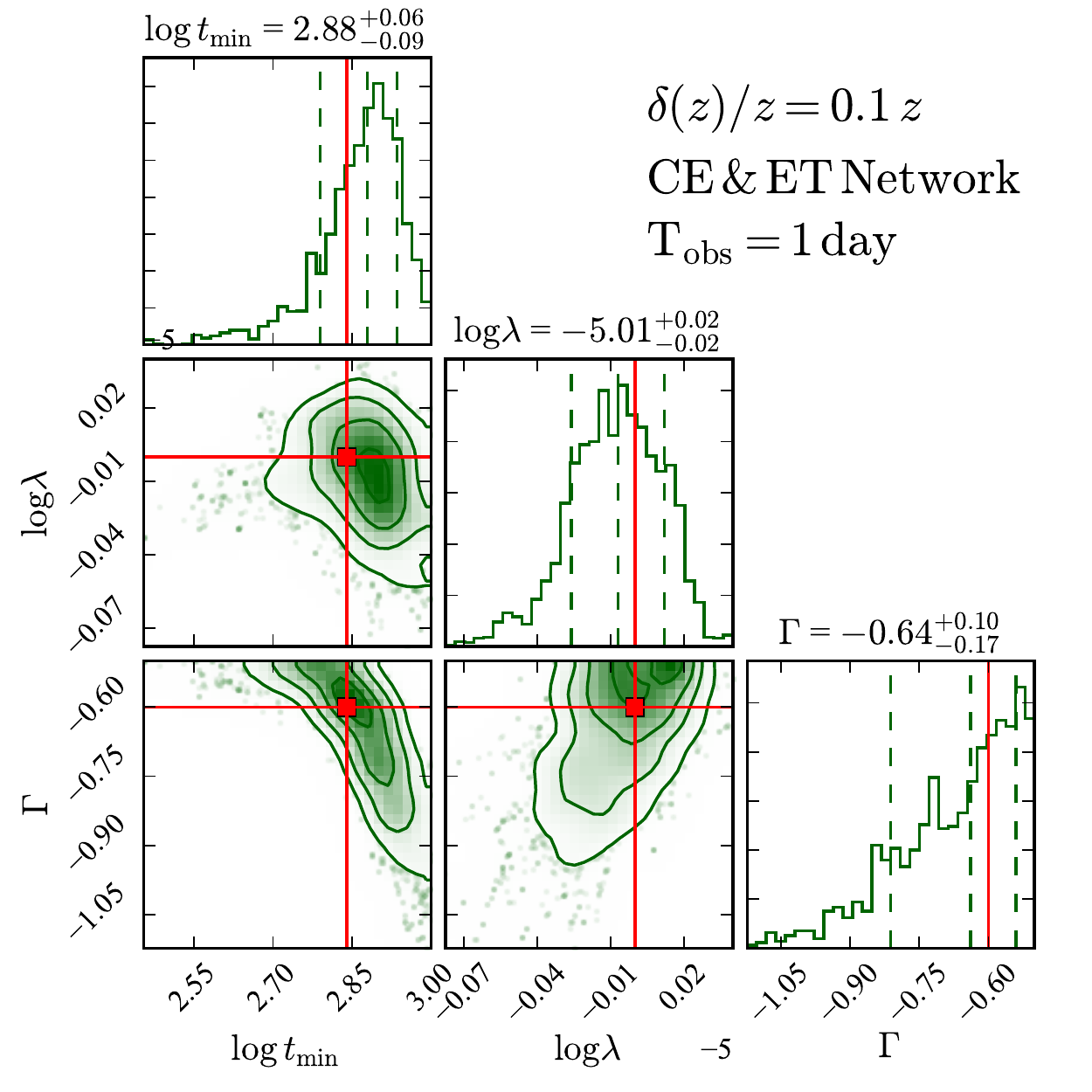}
\hspace{8pt}
\includegraphics[width=8cm]{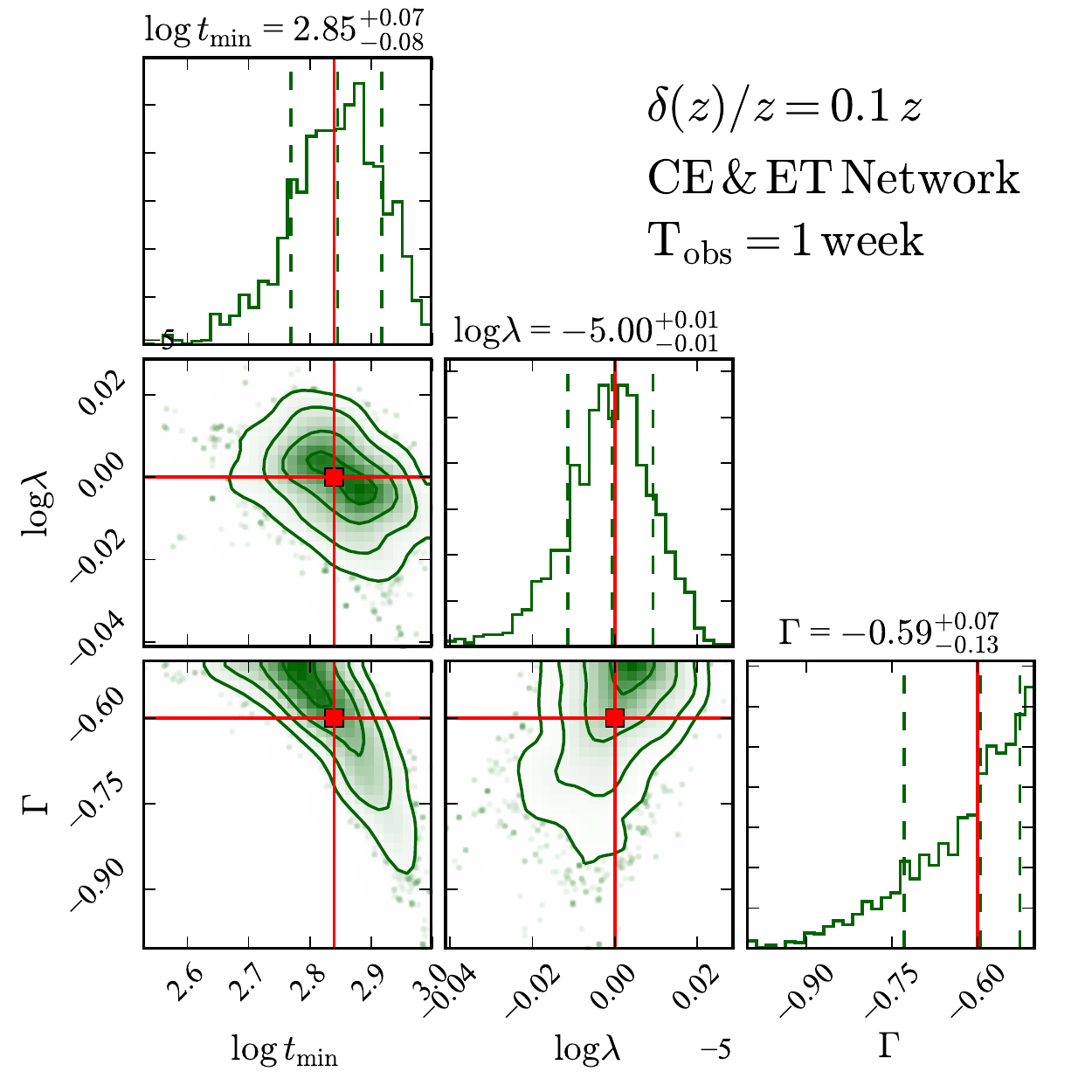}
\hspace{8pt}
\includegraphics[width=8cm]{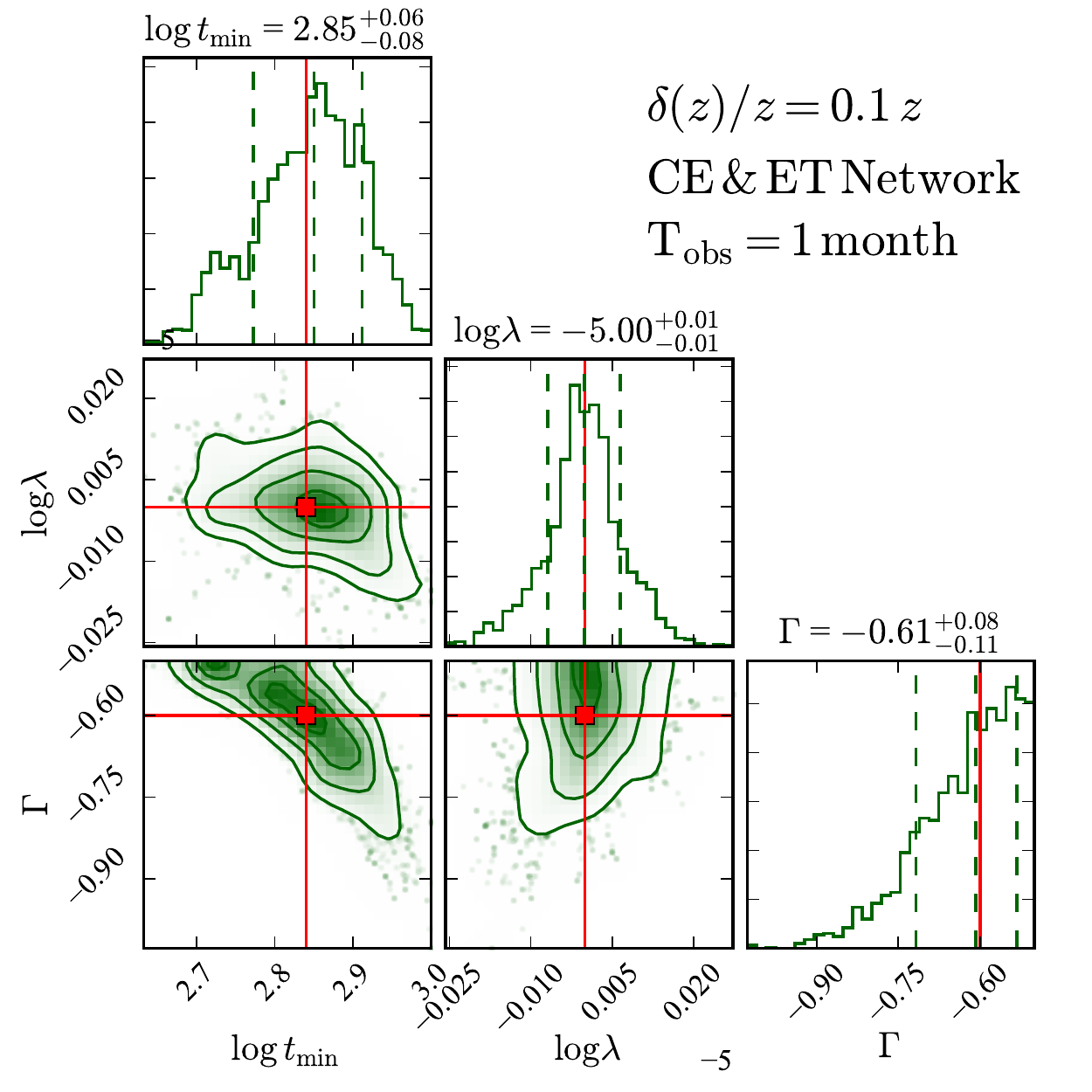}
\hspace{8pt}
\includegraphics[width=8cm]{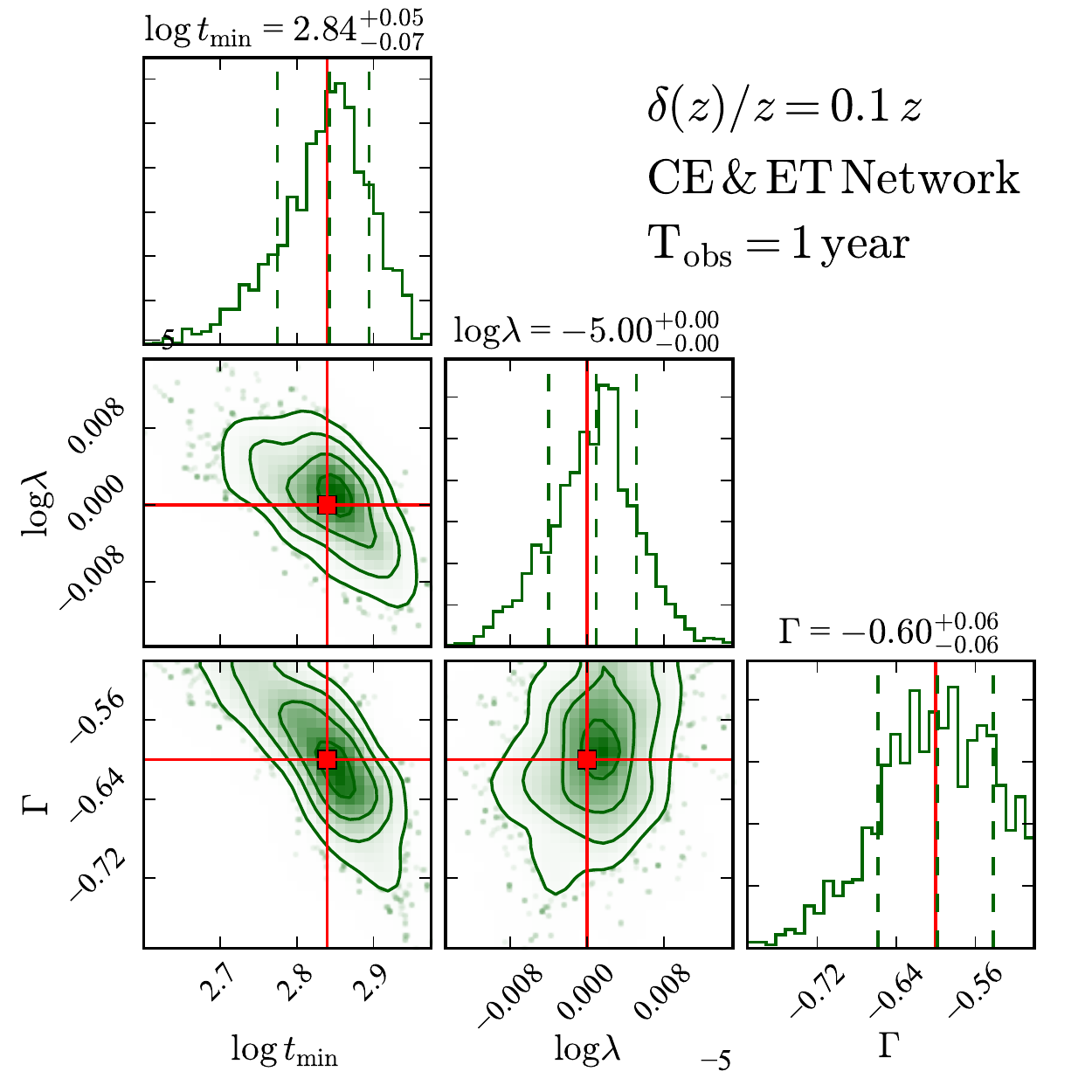}
\caption{Results of MCMC parameter estimation for a network of CE+ET with a range of operating timescales, spanning 1 day to 1 year.  The red vertical lines and circles mark the input DTD model, while the green curves and contours show the posteriors of the model parameters.  The black lines show the median and range of 16th to 84th percentiles.  In this case the redshift uncertainty is modeled as $\delta z/z=0.1z$.  With a year of observations all of the DTD parameters can be determined accurately to high precision.}
\label{f:network}
\end{figure*}

The predicted {\it observed} redshift distribution for the current generation of detectors at design sensitivity is shown in Figure~\ref{f:DRs}.  As expected, because the detection distance is limited to only a few hundred Mpc, all of the DTDs predict the same shape of observed distribution, with a simple change in scaling that can be accommodated by varying the unknown value of $\lambda$.

For the purpose of assessing the resulting constraints on the DTD and $\lambda$ we assume that BNS mergers from the current GW network will have precisely determined redshifts through associated EM counterparts and host galaxies.  Therefore, the error budget is dominated by the Poisson error based on the detection rate.  For the input model we assume $\Gamma=-0.6$, $t_{\rm min}=700$ Myr, and $\lambda=10^{-5}$ M$_\odot^{-1}$. Using our MCMC approach we show the resulting constraints on the DTD parameters for a year of Advanced LIGO/Virgo operations at design sensitivity in Figure~\ref{f:mcmc_adLIGO}. The results indicate that the DTD remains largely unconstrained, with the posterior distributions strongly influenced by the flat priors.  In particular, $\tmin$ is unconstrained, while $\Gamma$ and $\lambda$ show a strong degeneracy, with median values that are biased away from the injected model.  We find the same result for a decade of Advanced LIGO/Virgo operations.  

Our results for Advanced LIGO/Virgo suggest that even the proposed upgrades to the current facilities, such as A+ \citep{Miller:2014kma} and Voyager \citep{LIGOInstrumentWhitePaper} will not have a significant impact on the DTD since these facilities will still only detect BNS mergers in the local universe (see e.g., Figure 1, right panel of \citealt{Reitze:2019dyk}). As argued in Paper I, a more robust constraint on the DTD from the current generation of GW detectors may be achieved through the mass distribution of BNS merger host galaxies.  However, even this approach leaves a lingering degeneracy between $\Gamma$ and $\tmin$.

\section{results for Third-Generation Detectors}
\label{sec:3g}

The situation is drastically different for the third-generation detectors, ET and CE. In Figure~\ref{f:third_gen} we plot the expected detection rate as a function of redshift for ET, CE, and a network of ET+CE. Two improvements are readily apparent. First, the expected detection rate is about three orders of magnitude larger than for Advanced LIGO/Virgo.  Second, the redshift range for BNS merger detections increases to $z\sim 5$ in the case of ET+CE.  The latter improvement results in a clear difference between the redshift distributions of the various DTDs, while the former improvement provides the detection statistics needed to distinguish between the DTD models.  The differences between the various DTDs can no longer be scaled away with a change in $\lambda$ (as is the case for Advanced LIGO/Virgo). In what follows we focus on the case of ET+CE as a realistic version of a third-generation detector network.

In Figure~\ref{f:network} we show the result of MCMC fitting for the same input model used in the previous section.  We show the results for 1 \emph{day}, 1 \emph{week}, 1 \emph{month}, and 1 \emph{year} of observations. Unlike in the case of the current generation of GW detectors, we assume that the BNS mergers at cosmological distances will generally not have detectable EM counterparts (see Appendix~\ref{appendixC}.).  Instead we rely on distance information from the GW signal itself.  We model the resulting redshift uncertainty as $\delta z/z=0.1 z$; a detailed motivation for this parametrization is provided in Appendix~\ref{appendixB}. 

Our results show that an ET+CE network is able to constrain the DTD parameters and overcome the intrinsic degeneracy between $\Gamma$ and $\tmin$ within a year of observations.  The values of $\Gamma$ and $\tmin$ can be determined to better than $10\%$ accuracy. We note that these numbers depend on the overall event rate, which is determined by $\lambda$; here we use an injected value of $10^{-5}\msun^{-1}$, but the results can be rescaled for higher or lower values. 

\begin{figure*}
\setlength{\tabcolsep}{1mm}
%\hspace{0cm}
\includegraphics[width=\columnwidth]{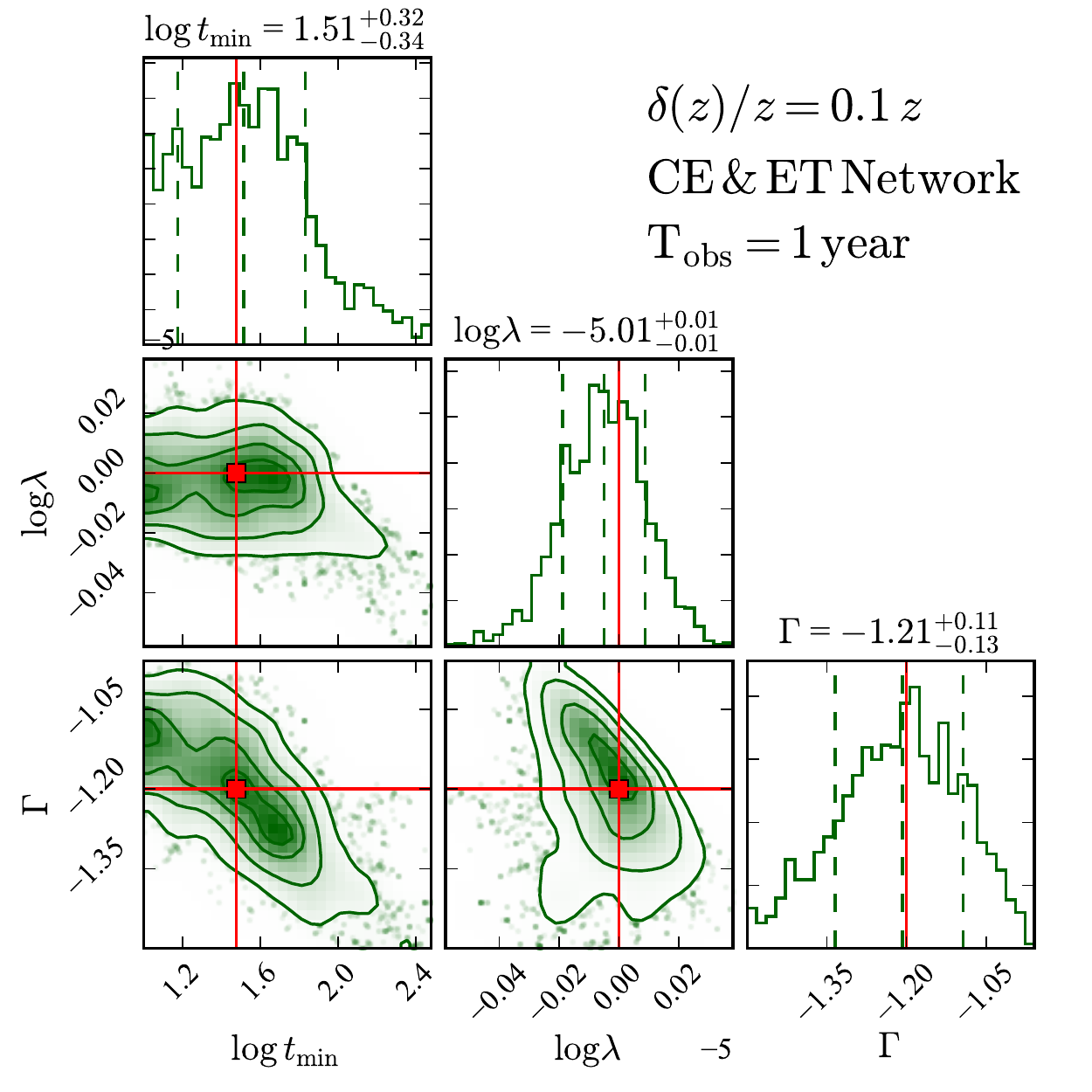}
\includegraphics[width=\columnwidth]{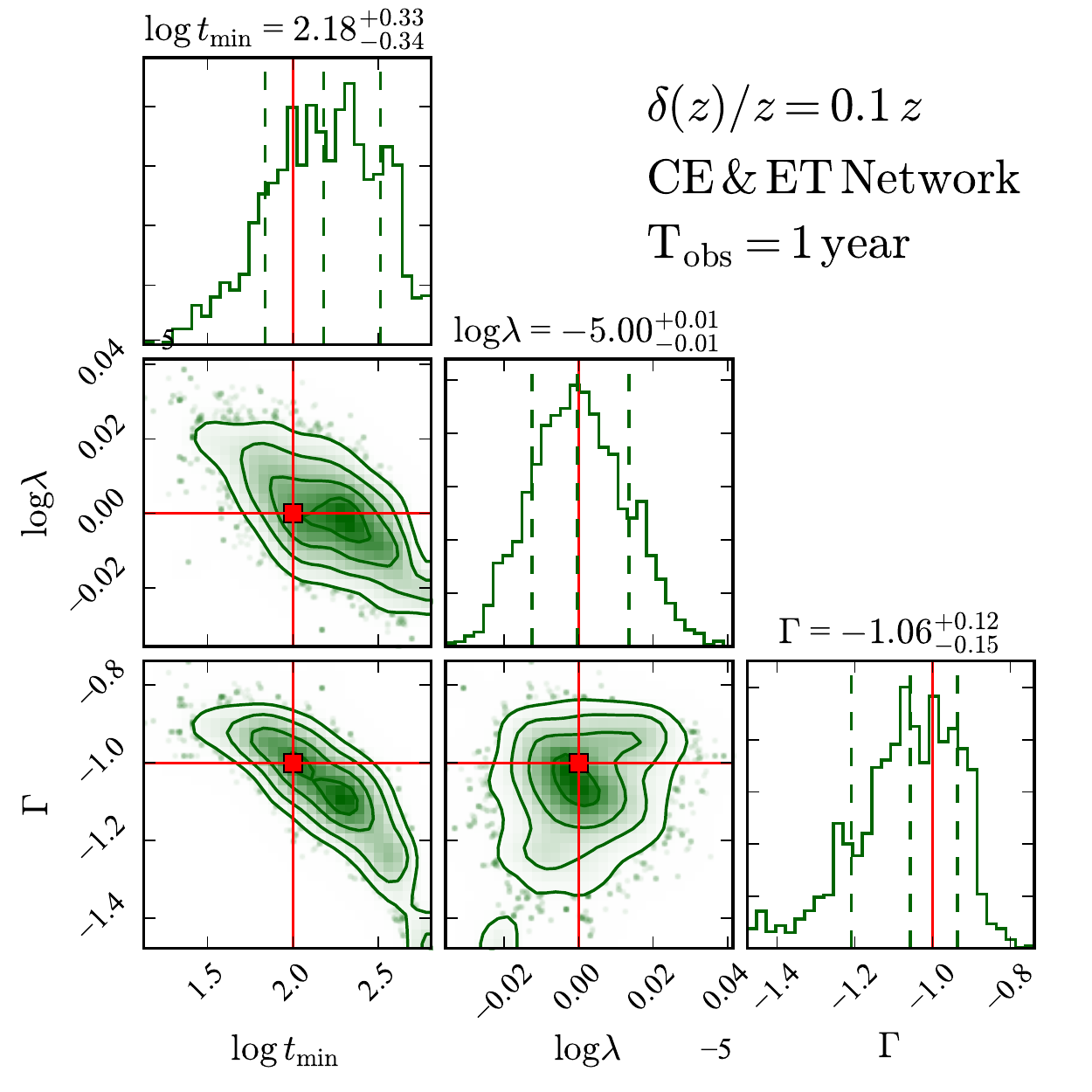}
\caption{ The same as in Figure~\ref{f:network} but for two DTD models that favor short merger timescales. \emph{Left:} An injected DTD with $\Gamma=-1.2$ and $t_{\rm min}=30$ Myr. \emph{Right:} An injected DTD with $\Gamma=-1.$ and $t_{\rm min}=100$ Myr. Although the value of $\Gamma$ is still recovered with about $10\%$ uncertainty in both cases, $t_{\rm min}$ becomes more challenging to accurately determine when its value is small.}
\label{f:network_steep_DTD}
\end{figure*}

To assess the impact of our input model on the results, in Figure~\ref{f:network_steep_DTD} we repeat the same exercise, but for two different DTDs that favor short merger timescales: $\Gamma=-1.2$ with $t_{\rm min}=30$ Myr, and $\Gamma=-1$ with $t_{\rm min}=100$ Myr. Although the power law index is recovered with the same accuracy as before, we find that $t_{\rm min}$ becomes more challenging to determine when its value is small. This is because the relative shift in the {\it observed} BNS merger redshift distribution becomes progressively smaller for small values of $\tmin$, which is challenging to detect in the presence of realistic GW redshift uncertainties.

\section{Summary and Discussion}
\label{sec:summary}

We investigated how well the DTD and mass efficiency of BNS systems can be determined through the redshift distribution of BNS mergers detected by current and future GW networks.  We model the DTD as a power law with a minimum merger timescale, and leave the mass efficiency as a free parameter.  While other DTDs have been proposed (e.g., \citealt{Simonetti:2019uq}), our primary conclusions should not be affected by the exact form of the DTD.  

We find that current GW detectors, which can only detect BNS mergers in the local universe, cannot directly constrain the DTD due to their limited sensitivity.  In effect, the various DTDs, with an appropriate scaling of $\lambda$, predict the same BNS merger detection rate at $z\approx 0$.  However, the situation is dramatically different for the anticipated third-generation detectors, which will be able to detect BNS mergers to $z\approx 5$. For this cosmological population, even in the presence of redshift uncertainties of $\delta z/z\approx 0.1z$ from the GW data, the large detection rate and broad redshift range will precisely determine the DTD within about a year of operations. It has been previously argued that the cosmological merger population uncovered by third-generation detectors will be able to constrain cosmological parameters \citep{Sathyaprakash:2009jc,Taylor:2011ex,Taylor:2012jo,VitaleFarr18}; our results for the DTD further bolster the science case for third-generation GW detectors.

\acknowledgements

This work was supported by the National Science Foundation under grant AST14-07835 and by NASA under theory grant NNX15AK82G. 
The Berger Time-Domain Group at
Harvard is supported in part by NSF under grant AST-1714498 and by
NASA under grant NNX15AE50G. MTS is thankful to Harvard-Smithsonian
Center for Astrophysics for hospitality which made this work possible.
SV, KKYN and CW acknowledge support of the National Science Foundation, the LIGO Laboratory and the LIGO
Data Grid clusters. LIGO was constructed by the California Institute of Technology and Massachusetts Institute of Technology with funding from the National Science Foundation and operates under cooperative agreement PHY-0757058.

\bibliographystyle{apj}
\bibliography{the_entire_lib}

\appendix
\chapter{}
\input{appendix_CTheta.tex}
\section{redshift uncertainty from GW detections}
\label{appendixB}

In the case of third-generation detectors it is unlikely that most BNS merger detections will have EM counterparts.  Instead, the distance information will need to be gleaned from the GW signal itself.  There is an inherent degeneracy between a binary's inclination angle in the sky with respect to a GW detector, $\theta_{\rm JN}$, and the luminosity distance \citep{Schutz:2011fn,Abbottetal:2016kd,Chen:2018us,Usman:2018tv}. \citet{Messenger:2012is} show that for a range of representative neutron star equations of state the redshift of such systems can be determined to an accuracy of $\sim 8-40\%$ for $z<1$ and $\sim 9-65\%$ for $1<z<4$. However, for a binary to be detectable at high redshifts, we expect the inclination to be close to face-on ($\theta_{\rm JN}=0^o$), or face-off ($\theta_{\rm JN}=180^o$), as most of the energy in gravitational waves is released along the angular momentum vector of the binary. \citet{Schutz:2011fn} derived an analytic formulation for the distribution of inclination angle of sources detectable by advanced detectors; $\lesssim 7\%$ ($\lesssim 3\%$) of detectable events will have viewing angles of $>70^o$ ($>80^o$) \citep{Chen:2018us}.

In principle, we can simulate BNS signals at different redshifts in third-generation detectors and estimate the evolution of uncertainty in parameter estimation using MCMC. Since BNS are a low-mass system with a long coalescence time, the parameter estimation is computationally expensive. To approximate the redshift uncertainty of a BNS at fixed redshift, we may extrapolate a binary black hole (BBH) signal by lowering the starting frequency from 10 Hz to 5 Hz, as to mimic the long inspiral phase in a BNS merger. We obtain the redshift uncertainties of BBHs in CE+ET from the simulations of \citet{2018PhRvD..98b4029V}, and fit the mean redshift uncertainties as a function of true redshifts. Then we calibrate our fit using the above extrapolation scheme to approximate the redshift uncertainties as $\delta z/z\approx 0.1z$.

Alternatively, \citet{Chen:2018us} developed a rapid algorithm that provides a luminosity distance uncertainty estimate for a large population of BNS merger detections. We use this algorithm to simulate 2000 BNS detections in a third-generation network, and compare the results to the extrapolation procedure above. We find that the two approaches yield consistent results. We therefore use the extrapolated distance uncertainty, and convert it to the redshift uncertainty through the adopted cosmology in this work. 

\section{redshift uncertainty from joint SGRB detections}
\label{appendixC}

The distance-inclination degeneracy can be broken through the detections of an associated SGRB. The relative fraction of on-axis mergers is only a few percent. A $\gamma$-ray detection alone will thereby reduce the overall redshift uncertainty of at most a few percent of BNS merger detections, and likely over a restricted redshift range (perhaps to $z\sim 2$).  An afterglow detection can further lead to a precise redshift determination through an associated host galaxy, but to date such detections have mainly been limited to $z\lesssim 1$ \citep{Berger14}, which is generally not high enough to make an impact on the DTD determination (see Figure~\ref{f:third_gen}). Thus, it seems unlikely that associated SGRBs and their afterglows will substantially improve the constraints on the DTD.

\end{document}

%% file: appendix_CTheta.tex
\section{Detection probability of a network of detectors}
\label{CofTheta}

The strain measured by a GW interferometer in frequency domain is given by
\begin{align}
\Tilde{h}(f)=F_+ \Tilde{h}_+(f) + F_{\times}\Tilde{h}_{\times}(f),
\end{align}
where $\Tilde{h}_{+,\times}$ are the $+,\times$-polarization bases and $F_{+,\times}$ are the corresponding beam pattern functions,
\begin{align}
F_+&=g\left[\frac{1}{2}\left( 1+\cos^2{\theta} \right) \cos{2\phi}\cos{2\psi}-\cos{\theta}\sin{2\phi}\sin{2\psi} \right],\\
F_{\times}&=g \left[\frac{1}{2}\left( 1+\cos^2{\theta} \right) \cos{2\phi}\sin{2\psi}+\cos{\theta}\sin{2\phi}\cos{2\psi} \right],
\end{align}
where $\theta,\phi$ and $\psi$ are the zenith, azimuth and polarization angles respectively, and $g$ is a dimensionless coefficient determined by the geometry of an interferometer~\citep{sathya:gwreview,Chen:gw_dist,schutz:antenna}.
CE is a single interferometer with the angle between two arms equal to $90\deg$, hence $g_{\text{CE}}=1$~\citep{Chen:gw_dist,schutz:antenna}.
ET consists of 3 identical interferometers with the angle between two arms equal to $60\deg$, forming an equilateral triangle, hence $g_{\text{ET}}=\sqrt{3}/2$ for each interferometer in ET~\citep{sathya:ETcqg,Punturo:2010jf}.

The detection probability, $P_{\text{det}}$, is defined as the probability of a detection with $\rho_{\text{net}}\geq \rho_{T}$, where $\rho_{T}$ is the SNR threshold of detection and $\rho^2_{\text{net}}=\sum_i{\rho^2_i}$ is the network SNR as a geometric sum of SNR of each inteferometer.
Assuming isotropic sky locations, orbital orientation and polarization, (i.e., uniform distribution of $(\cos\theta, \phi, \psi, \cos\iota)$), $P_{\text{det}}$ is given analytically as
\begin{align}\label{pdet}
    P_{\text{det}}\left( \theta_{\text{int}} , z \right) = \int H\left( \frac{\rho_{T}}{ \rho_{\text{net}}(\theta_{\text{int}}, z, \theta, \phi, \psi, \iota)} \right) d\cos{\theta} d\phi d\psi d\cos{\iota} ,
\end{align}
where $\theta_{\text{int}}$ is the set of intrinsic parameters, which are fixed at $1.4-1.4\msun$ and zero-spin for BNS systems, and $H(w(z,\theta, \phi, \psi, \iota))$ is the unitary step function defined in $w\in(0,1]$ for $w(z, \theta, \phi, \psi, \iota)=\rho_T/\rho_{\text{net}}(z, \theta, \phi, \psi, \iota)$.

For single CE or ET, we follow the inspiral approximation in~\cite{Finn96}. The SNR of a single interferometer is approximately
\begin{align}
\rho^2=64 \Theta^2 \left(\frac{r_0}{d_L}\right)^2 \left(\frac{\mathcal{M}_z}{1.2\msun}\right)^{5/3} \zeta^2(f_{\text{max}}),
\end{align}
where $\Theta^2=4\left[ F_+^2\left(1+\cos^2\iota\right)^2 + 4F_{\times}^2\cos^2\iota\right]$, and proportional to $\Theta^2_{\text{net}}=\sum_i{\Theta^2_i}$ for the same signal strain observed by homogeneous detectors such as single CE or ET.
Hence $P_{\text{det}}(w(z))$ is equivalently the survival function of $\Theta_{\text{net}}/\Theta^{\text{max}}_{\text{net}}$, where $\Theta^{\text{max}}_{\text{net}}$ is the maximum response of a particular network of detectors \citep{sathya:gwreview,Chen:gw_dist,schutz:antenna,Finn96}.
Here $w(z)=\rho_T/\rho_{\text{opt}}(z)$ where $\rho_{\text{opt}}$ is the optimal SNR at maximum $\Theta^2$.
CE has $\Theta^{\text{max}}_{\text{CE}}=4$ and ET has $\Theta^{\text{max}}_{\text{ET}}=4\times\sqrt{3\times3/4}=6$.
We have verified that the inspiral approximation in 3G detectors only results in few-percent difference in SNR for $z\lesssim 6$, which is the region we are interested in.

Approximated forms of $P_{\text{det}}$ assuming uniformly distributed $(\cos\theta, \phi, \psi, \cos\iota)$ provided in \citet{Finn96} or \citet{Dominik:2015dp} are only suitable for a single CE or a second-generation network. Therefore, we generate the survival function of $\Theta$ by drawing $10^6$ points of uniformly distributed $(\cos\theta, \phi, \psi, \cos\iota)$ in single CE or ET and fit the survival function with the following parametric form:
\begin{align}\label{eq:erf_fit}
P_{\text{det}}(w;A,B)=\frac{\text{erf}\left( A-Bw \right) - \text{erf}\left( A-B \right)}{\text{erf}\left( A \right) - \text{erf}\left( A-B \right)},
\end{align}
where $\text{erf}(x)=\frac{2}{\sqrt{\pi}}\int_0^x e^{-t^2}dt$ is the error function. We also employ a 10th order polynomial,
\begin{align}\label{eq:polyfit}
P_{\text{det}}(w;{a_k})=\sum_{k=1}^9 a_k (1-w)^k+\left(1-\sum_{k=1}^9 a_k\right)\left(1-w\right)^{10},
\end{align}
where $\{a_k\}$ are the polynomial coefficients.

The above simplification using distribution of $\Theta$ breaks down for a CE+ET network due to the different sensitivity and heterogeneous geometry of each interferometer. Instead, we calculate the integral in Equation~\ref{pdet} by simulating waveform and network SNR for each redshift. Then $w(z)=z/z_{\text{horizon}}$ and $z_{\text{horizon}}$ is the redshift of the detector horizon, which is $\sim 12.5$ in CE+ET for a BNS merger.  Again we fit the simulation result for CE+ET  using Equations~\ref{eq:erf_fit} and \ref{eq:polyfit}.

\begin{figure}
    \centering
    \includegraphics[width=0.7\textwidth]{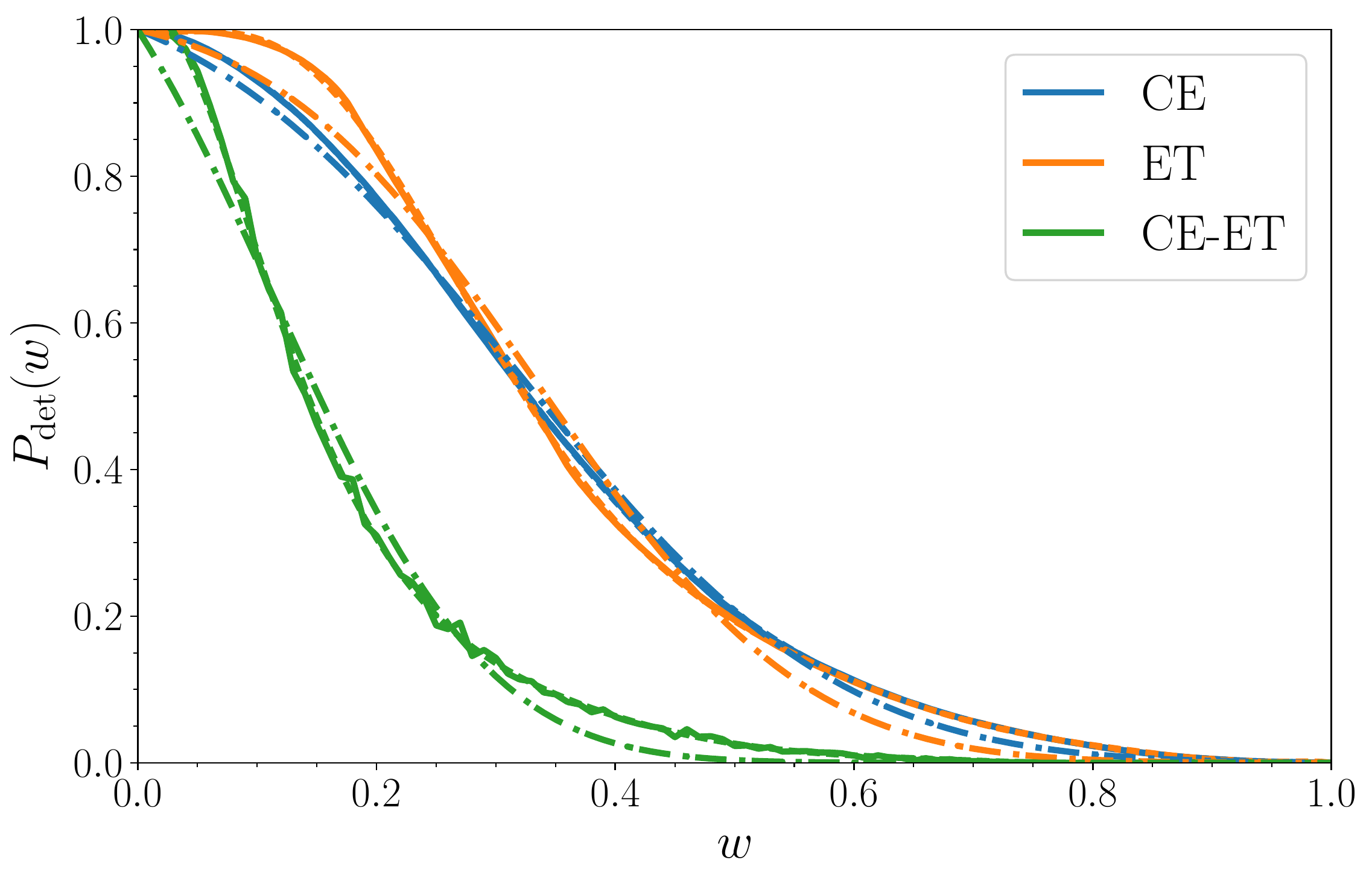}
    \caption{Comparison of simulation and parametric fits for CE, ET and CE+ET network. Solid lines show the simulation, dashed-dotted lines show the fit using Equation~\ref{eq:erf_fit} and dashed lines show the fit using Equation~\ref{eq:polyfit}. Here $w=\rho_T/\rho_{\text{opt}}(z)$ for single CE or ET and $w=z/z_{\text{horizon}}$ for CE+ET network.}
    \label{fig:p_det_fit}
\end{figure}

Tables~\ref{tab:erf_fit} and \ref{tab:polyfit} show the definitions of $w$ and fitting parameters of Equations~\ref{eq:erf_fit} and \ref{eq:polyfit} in each network. Figure~\ref{fig:p_det_fit} shows the comparison of actual simulation and the parametric fits.

\begin{table}[!ht]
\centering
\begin{tabular}{l|l|l|l}
Network& $w(z)$ & $A$ & $B$ \\
\hline
CE & $\rho_T/\rho_{\text{opt}}(z)$ & 1.63 & 3.93 \\
ET & $\rho_T/\rho_{\text{opt}}(z)$ & 1.05 & 3.33 \\
CE-ET & $z/z_{\text{horizon}}$ & -0.58 & -5.05
\end{tabular}
\caption{Definition of $w$ and fitting parameters of Equation~\ref{eq:erf_fit} for CE, ET and CE+ET.\label{tab:erf_fit}}
\end{table}

\begin{table}[!ht]
\centering
\begin{tabular}{l|l|l|l|l|l|l|l|l|l|l}
 Network& $w(z)$ & $a_1$ & $a_2$ & $a_3$ & $a_4$ & $a_5$ & $a_6$ & $a_7$ & $a_8$ & $a_9$ \\
 \hline
 CE    & $\rho_T/\rho_{\text{opt}}(z)$ &   0.038    &   -0.57    &   10.76    &   -49.76    &   143.1    &   -293.1    &  470.5     &   -512.4    &   307.5    \\
ET    & $\rho_T/\rho_{\text{opt}}(z)$&  0.22     &   -8.51    &   132.8    &    -966.0   &   3995.7    &  -9923.5     &        15017.1  & -13487.3  &   6581.3    \\
CE-ET &  $z/z_{\text{horizon}}$ & -0.15    &  5.48     &   -71.4    &   455.0    &   -1613.3    &   3363.5    & -4119.8      &   2794.4    & -877.2
\end{tabular}
\caption{Definition of $w$ and fitting parameters of Equation~\ref{eq:polyfit} for CE, ET and CE+ET.\label{tab:polyfit}}
\end{table}

%% file: DTD.bbl
\begin{thebibliography}{}
\expandafter\ifx\csname natexlab\endcsname\relax\def\natexlab#1{#1}\fi

\bibitem[{Abbott {et~al.}(2017{\natexlab{a}})Abbott, Abbott, Abbott, Abernathy,
  Ackley, Adams, Addesso, Adhikari, Adya, Affeldt, Aggarwal, Aguiar, Ain,
  Ajith, Allen, Altin, Anderson, Anderson, Arai, Araya, Arceneaux, Areeda,
  Arun, Ashton, Ast, Aston, Aufmuth, Aulbert, Babak, Baker, Ballmer, Barayoga,
  Barclay, Barish, Barker, Barr, Barsotti, Bartlett, Bartos, Bassiri, Batch,
  Baune, Bell, Berger, Bergmann, Berry, Betzwieser, Bhagwat, Bhandare, Bilenko,
  Billingsley, Birch, Birney, Biscans, Bisht, Biwer, Blackburn, Blair, Blair,
  Blair, Bock, Bogan, Bohe, Bond, Bork, Bose, Brady, Braginsky, Brau,
  Brinkmann, Brockill, Broida, Brooks, Brown, Brown, Brown, Brunett, Buchanan,
  Buikema, Buonanno, Byer, Cabero, Cadonati, Cahillane, Calder{\'o}n~Bustillo,
  Callister, Camp, Cannon, Cao, Capano, Caride, Caudill, Cavagli{\`a}, Cepeda,
  Chamberlin, Chan, Chao, Charlton, Cheeseboro, Chen, Chen, Cheng, Cho, Cho,
  Chow, Christensen, Chu, Chung, Ciani, Clara, Clark, Collette, Cominsky,
  Constancio~Jr, Cook, Corbitt, Cornish, Corsi, Costa, Coughlin, Coughlin,
  Countryman, Couvares, Cowan, Coward, Cowart, Coyne, Coyne, Craig, Creighton,
  Cripe, Crowder, Cumming, Cunningham, Dal~Canton, Danilishin, Danzmann,
  Darman, Dasgupta, Da~Silva~Costa, Dave, Davies, Daw, De, DeBra, Del~Pozzo,
  Denker, Dent, Dergachev, DeRosa, DeSalvo, Devine, Dhurandhar, D{\'\i}az,
  Di~Palma, Donovan, Dooley, Doravari, Douglas, Downes, Drago, Drever,
  Driggers, Dwyer, Edo, Edwards, Effler, Eggenstein, Ehrens, Eichholz,
  Eikenberry, Engels, Essick, Etzel, Evans, Evans, Everett, Factourovich, Fair,
  Fairhurst, Fan, Fang, Farr, Farr, Favata, Fays, Fehrmann, Fejer, Fenyvesi,
  Ferreira, Fisher, Fletcher, Frei, Freise, Frey, Fritschel, Frolov, Fulda,
  Fyffe, Gabbard, Gair, Gaonkar, Gaur, Gehrels, Geng, George, Gergely, Ghosh,
  Ghosh, Giaime, Giardina, Gill, Glaefke, Goetz, Goetz, Gondan, Gonz{\'a}lez,
  Gopakumar, Gordon, Gorodetsky, Gossan, Graef, Graff, Grant, Gras, Gray,
  Green, Grote, Grunewald, Guo, Gupta, Gupta, Gushwa, Gustafson, Gustafson,
  Hacker, Hall, Hall, Hammond, Haney, Hanke, Hanks, Hanna, Hannam, Hanson,
  Hardwick, Harry, Harry, Hart, Hartman, Haster, Haughian, Heintze, Hendry,
  Heng, Hennig, Henry, Heptonstall, Heurs, Hild, Hoak, Holt, Holz, Hopkins,
  Hough, Houston, Howell, Hu, Huang, Huerta, Hughey, Husa, Huttner, Huynh-Dinh,
  Indik, Ingram, Inta, Isa, Isi, Isogai, Iyer, Izumi, Jang, Jani, Jawahar,
  Jian, Jim{\'e}nez-Forteza, Johnson, Jones, Jones, Ju, Haris, Kalaghatgi,
  Kalogera, Kandhasamy, Kang, Kanner, Kapadia, Karki, Karvinen, Kasprzack,
  Katsavounidis, Katzman, Kaufer, Kaur, Kawabe, Kehl, Keitel, Kelley, Kells,
  Kennedy, Key, Khalili, \& Khan}]{Abbott:2017ie}
Abbott, B.~P., Abbott, R., Abbott, T.~D., {et~al.} 2017{\natexlab{a}},
  Classical and Quantum Gravity, 34, 044001

\bibitem[{Abbott {et~al.}(2017{\natexlab{b}})Abbott, Abbott, Abbott, Acernese,
  Adams, Adams, Addesso, Adhikari, Adya, Affeldt, Afrough, Agarwal, Agathos,
  Agatsuma, Aggarwal, Aguiar, Aiello, Ain, Ajith, Allen, Allen, Allocca, Altin,
  Amato, Ananyeva, Anderson, Anderson, Angelova, Antier, Appert, Arai, Araya,
  Areeda, Arnaud, Arun, Ascenzi, Ashton, Ast, Aston, Astone, Atallah, Aufmuth,
  Aulbert, AultONeal, Austin, Avila-Alvarez, Babak, Bacon, Bader, Bae, Baker,
  Baldaccini, Ballardin, Ballmer, Banagiri, Barayoga, Barclay, Barish, Barker,
  Barkett, Barone, Barr, Barsotti, Barsuglia, Barta, Barthelmy, Bartlett,
  Bartos, Bassiri, Basti, Batch, Bawaj, Bayley, Bazzan, B{\'e}csy, Beer,
  Bejger, Belahcene, Bell, Berger, Bergmann, Bero, Berry, Bersanetti,
  Bertolini, Betzwieser, Bhagwat, Bhandare, Bilenko, Billingsley, Billman,
  Birch, Birney, Birnholtz, Biscans, Biscoveanu, Bisht, Bitossi, Biwer,
  Bizouard, Blackburn, Blackman, Blair, Blair, Blair, Bloemen, Bock, Bode,
  Boer, Bogaert, Bohe, Bondu, Bonilla, Bonnand, Boom, Bork, Boschi, Bose,
  Bossie, Bouffanais, Bozzi, Bradaschia, Brady, Branchesi, Brau, Briant,
  Brillet, Brinkmann, Brisson, Brockill, Broida, Brooks, Brown, Brown, Brunett,
  Buchanan, Buikema, Bulten, Buonanno, Buskulic, Buy, Byer, Cabero, Cadonati,
  Cagnoli, Cahillane, Bustillo, Callister, Calloni, Camp, Canepa, Canizares,
  Cannon, Cao, Cao, Capano, Capocasa, Carbognani, Caride, Carney, Diaz,
  Casentini, Caudill, Cavagli{\`a}, Cavalier, Cavalieri, Cella, Cepeda,
  Cerd{\'a}-Dur{\'a}n, Cerretani, Cesarini, Chamberlin, Chan, Chao, Charlton,
  Chase, Chassande-Mottin, Chatziioannou, Cheeseboro, Chen, Chen, Chen, Cheng,
  Chia, Chincarini, Chiummo, Chmiel, Cho, Cho, Chow, Christensen, Chu, Chua,
  Chua, Chung, Chung, Ciani, Ciolfi, Cirelli, Cirone, Clara, Clark, Clearwater,
  Cleva, Cocchieri, Coccia, Cohadon, Cohen, Colla, Collette, Cominsky, Jr,
  Conti, Cooper, Corban, Corbitt, Cordero-Carri{\'o}n, Corley, Cornish, Corsi,
  Cortese, Costa, Coughlin, Coughlin, Coulon, Countryman, Couvares, Covas,
  Cowan, Coward, Cowart, Coyne, Coyne, Creighton, Creighton, Cripe, Crowder,
  Cullen, Cumming, Cunningham, Cuoco, Canton, D{\'a}lya, Danilishin, D'Antonio,
  Danzmann, Dasgupta, Da~Silva~Costa, Dattilo, Dave, Davier, Davis, Daw, Day,
  De, DeBra, Degallaix, Laurentis, Del{\'e}glise, Pozzo, Demos, Denker, Dent,
  Pietri, Dergachev, Rosa, DeRosa, Rossi, DeSalvo, Varona, Devenson,
  Dhurandhar, D{\'\i}az, Fiore, Giovanni, Girolamo, Lieto, Pace, Palma, Renzo,
  Doctor, Dolique, Donovan, Dooley, Doravari, Dorrington, Douglas, {\'A}lvarez,
  Downes, Drago, Dreissigacker, Driggers, Du, Ducrot, Dupej, Dwyer, Edo,
  Edwards, Effler, Ehrens, Eichholz, Eikenberry, Eisenstein, Essick, Estevez,
  \& Et...}]{Abbott:2017it}
---. 2017{\natexlab{b}}, The Astrophysical Journal, 848, L12

\bibitem[{Abbott et~al .(2017)}]{Collaboration:2017kt}
Abbott et~al ., B.~P. 2017, arXiv.org, 161101

\bibitem[{Abbott et~al . {et~al.}(2016)Abbott et~al ., Collaboration, Abbott,
  Abbott, Abbott, Abernathy, Acernese, Ackley, Adams, Adams, Addesso, Adhikari,
  Adya, Affeldt, Agathos, Agatsuma, Aggarwal, Aguiar, Aiello, Ain, Ajith,
  Allen, Allocca, Altin, Anderson, Anderson, Arai, Araya, Arceneaux, Areeda,
  Arnaud, Arun, Ascenzi, Ashton, Ast, Aston, Astone, Aufmuth, Aulbert, Babak,
  Bacon, Bader, Baker, Baldaccini, Ballardin, Ballmer, Barayoga, Barclay,
  Barish, Barker, Barone, Barr, Barsotti, Barsuglia, Barta, Bartlett, Bartos,
  Bassiri, Basti, Batch, Baune, Bavigadda, Bazzan, Behnke, Bejger, Bell, Bell,
  Berger, Bergman, Bergmann, Berry, Bersanetti, Bertolini, Betzwieser, Bhagwat,
  Bhandare, Bilenko, Billingsley, Birch, Birney, Birnholtz, Biscans, Bisht,
  Bitossi, Biwer, Bizouard, Blackburn, Blair, Blair, Blair, Bloemen, Bock,
  Bodiya, Boer, Bogaert, Bogan, Bohe, \& Bojtos}]{Abbottetal:2016kd}
Abbott et~al ., B.~P., Collaboration, t.~V., Abbott, B.~P., {et~al.} 2016,
  Physical Review Letters, 688

\bibitem[{Behroozi {et~al.}(2014)Behroozi, Ramirez-Ruiz, \&
  Fryer}]{Behroozi:2014bp}
Behroozi, P.~S., Ramirez-Ruiz, E., \& Fryer, C.~L. 2014, The Astrophysical
  Journal, 792, 123

\bibitem[{Belczynski {et~al.}(2018)Belczynski, Bulik, Olejak, Chruslinska,
  Singh, Pol, Zdunik, O{\textquoteright}Shaughnessy, McLaughlin, Lorimer,
  Korobkin, Heuvel, Davies, \& Holz}]{Belczynski:2018vr}
Belczynski, K., Bulik, T., Olejak, A., {et~al.} 2018, 1812.10065

\bibitem[{{Berger}(2014)}]{Berger14}
{Berger}, E. 2014, \araa, 52, 43

\bibitem[{{Chen} {et~al.}(2017){Chen}, {Holz}, {Miller}, {Evans}, {Vitale}, \&
  {Creighton}}]{Chen:gw_dist}
{Chen}, H.-Y., {Holz}, D.~E., {Miller}, J., {et~al.} 2017, arXiv e-prints,
  arXiv:1709.08079

\bibitem[{Chen {et~al.}(2018)Chen, Vitale, \& Narayan}]{Chen:2018us}
Chen, H.-Y., Vitale, S., \& Narayan, R. 2018, 1807.05226

\bibitem[{C{\^o}t{\'e} {et~al.}(2018)C{\^o}t{\'e}, Fryer, Belczynski, Korobkin,
  Chruslinska, Vassh, Mumpower, Lippuner, Sprouse, Surman, \&
  Wollaeger}]{Cote:2018gj}
C{\^o}t{\'e}, B., Fryer, C.~L., Belczynski, K., {et~al.} 2018, The
  Astrophysical Journal, 855, 99

\bibitem[{Dominik {et~al.}(2012)Dominik, Belczynski, Fryer, Holz, Berti, Bulik,
  Mandel, \& O'Shaughnessy}]{Dominik:2012cwa}
Dominik, M., Belczynski, K., Fryer, C., {et~al.} 2012, The Astrophysical
  Journal, 759, 52

\bibitem[{Dominik {et~al.}(2015)Dominik, Berti, O'Shaughnessy, Mandel,
  Belczynski, Fryer, Holz, Bulik, \& Pannarale}]{Dominik:2015dp}
Dominik, M., Berti, E., O'Shaughnessy, R., {et~al.} 2015, The Astrophysical
  Journal, 806, 263

\bibitem[{{Finn}(1996)}]{Finn96}
{Finn}, L.~S. 1996, \prd, 53, 2878

\bibitem[{{Fong} {et~al.}(2015){Fong}, {Berger}, {Margutti}, \&
  {Zauderer}}]{fbm+15}
{Fong}, W., {Berger}, E., {Margutti}, R., \& {Zauderer}, B.~A. 2015, \apj, 815,
  102

\bibitem[{{Fong} {et~al.}(2013){Fong}, {Berger}, {Chornock}, {Margutti},
  {Levan}, {Tanvir}, {Tunnicliffe}, {Czekala}, {Fox}, {Perley}, {Cenko},
  {Zauderer}, {Laskar}, {Persson}, {Monson}, {Kelson}, {Birk}, {Murphy},
  {Servillat}, \& {Anglada}}]{fbc+13}
{Fong}, W., {Berger}, E., {Chornock}, R., {et~al.} 2013, \apj, 769, 56

\bibitem[{Foreman-Mackey {et~al.}(2013)Foreman-Mackey, Hogg, Lang, \&
  Goodman}]{ForemanMackey:2013io}
Foreman-Mackey, D., Hogg, D.~W., Lang, D., \& Goodman, J. 2013, Publications of
  the Astronomical Society of the Pacific, 125, 306

\bibitem[{{Hotokezaka} {et~al.}(2018){Hotokezaka}, {Beniamini}, \&
  {Piran}}]{Hotokezaka2018IJMPD}
{Hotokezaka}, K., {Beniamini}, P., \& {Piran}, T. 2018, International Journal
  of Modern Physics D, 27, 1842005

\bibitem[{Kim {et~al.}(2015)Kim, Perera, \& McLaughlin}]{Kim:2015bi}
Kim, C., Perera, B. B.~P., \& McLaughlin, M.~A. 2015, Monthly Notices of the
  Royal Astronomical Society, 448, 928

\bibitem[{{Komiya} {et~al.}(2014){Komiya}, {Yamada}, {Suda}, \&
  {Fujimoto}}]{Komiya:2014ie}
{Komiya}, Y., {Yamada}, S., {Suda}, T., \& {Fujimoto}, M.~Y. 2014, \apj, 783,
  132

\bibitem[{{Lantz} {et~al.}(2018){Lantz}, {Danilishin}, {Hild}, {Gustafson},
  {Coyne}, V., {Hammond}, {Adhikari}, {Evans}, \&
  R.}]{LIGOInstrumentWhitePaper}
{Lantz}, B., {Danilishin}, S., {Hild}, S., {et~al.} 2018

\bibitem[{{Leibler} \& {Berger}(2010)}]{lb10}
{Leibler}, C.~N., \& {Berger}, E. 2010, \apj, 725, 1202

\bibitem[{Madau \& Dickinson(2014)}]{Madau:2014gtb}
Madau, P., \& Dickinson, M. 2014, Annual Review of Astronomy and Astrophysics,
  52, 415

\bibitem[{Matteucci {et~al.}(2014)Matteucci, Romano, Arcones, Korobkin, \&
  Rosswog}]{Matteucci:2014jta}
Matteucci, F., Romano, D., Arcones, A., Korobkin, O., \& Rosswog, S. 2014,
  Monthly Notices of the Royal Astronomical Society, 438, 2177

\bibitem[{Messenger \& Read(2012)}]{Messenger:2012is}
Messenger, C., \& Read, J. 2012, Physical Review Letters, 108, 301

\bibitem[{Miller {et~al.}(2015)Miller, Barsotti, Vitale, Fritschel, Evans, \&
  Sigg}]{Miller:2014kma}
Miller, J., Barsotti, L., Vitale, S., {et~al.} 2015, Phys. Rev., D91, 062005

\bibitem[{O{\textquoteright}Shaughnessy
  {et~al.}(2008)O{\textquoteright}Shaughnessy, Belczynski, \&
  Kalogera}]{OShaughnessy:2008bm}
O{\textquoteright}Shaughnessy, R., Belczynski, K., \& Kalogera, V. 2008, The
  Astrophysical Journal, 675, 566

\bibitem[{Planck~Collaboration {et~al.}(2016)Planck~Collaboration, Ade,
  Aghanim, Arnaud, Ashdown, Aumont, Baccigalupi, Banday, Barreiro, Bartlett,
  Bartolo, Battaner, Battye, Benabed, Benoit, Benoit-Levy, Bernard, Bersanelli,
  Bielewicz, Bock, Bonaldi, Bonavera, Bond, Borrill, Bouchet, Boulanger,
  Bucher, Burigana, Butler, Calabrese, Cardoso, Catalano, Challinor, Chamballu,
  Chary, Chiang, Chluba, Christensen, Church, Clements, Colombi, {Colombo, L.
  P. L.}, Combet, Coulais, Crill, Curto, Cuttaia, Danese, Davies, Davis,
  de~Bernardis, de~Rosa, De~Zotti, Delabrouille, Desert, Di~Valentino,
  Dickinson, Diego, Dolag, Dole, Donzelli, Dor{\'e}, Douspis, Ducout, Dunkley,
  Dupac, Efstathiou, Elsner, Ensslin, Eriksen, Farhang, Fergusson, Finelli,
  Forni, Frailis, Fraisse, Franceschi, Frejsel, Galeotta, Galli, Ganga,
  Gauthier, Gerbino, Ghosh, Giard, Giraud-Heraud, Giusarma, Gjerlow,
  Gonz{\`a}lez-Nuevo, Gorski, Gratton, Gregorio, Gruppuso, Gudmundsson, Hamann,
  Hansen, Hanson, Harrison, Helou, Henrot-Versille, Hernandez-Monteagudo,
  Herranz, Hildebrandt, Hivon, Hobson, Holmes, Hornstrup, Hovest, Huang,
  Huffenberger, Hurier, Jaffe, Jaffe, Jones, Juvela, Keihanen, Keskitalo,
  Kisner, Kneissl, Knoche, Knox, Kunz, Kurki-Suonio, Lagache, Lahteenmaki,
  Lamarre, Lasenby, Lattanzi, Lawrence, Leahy, Leonardi, Lesgourgues, Levrier,
  Lewis, Liguori, Lilje, Linden-Vornle, Lopez-Caniego, Lubin, Macias-Perez,
  Maggio, Maino, Mandolesi, Mangilli, Marchini, Maris, Martin, Martinelli,
  Martinez-Gonzalez, Masi, Matarrese, McGehee, Meinhold, Melchiorri, Melin,
  Mendes, Mennella, Migliaccio, Millea, Mitra, Miville-Deschenes, Moneti,
  Montier, Morgante, Mortlock, Moss, Munshi, Murphy, Naselsky, Nati, Natoli,
  Netterfield, Norgaard-Nielsen, Noviello, Novikov, Novikov, Oxborrow, Paci,
  Pagano, Pajot, Paladini, Paoletti, Partridge, Pasian, Patanchon, Pearson,
  Perdereau, Perotto, Perrotta, Pettorino, Piacentini, Piat, Pierpaoli,
  Pietrobon, Plaszczynski, Pointecouteau, Polenta, Popa, Pratt, Prezeau,
  Prunet, Puget, Rachen, Reach, Rebolo, Reinecke, Remazeilles, Renault, Renzi,
  Ristorcelli, Rocha, Rosset, Rossetti, Roudier, Rouille~d'Orfeuil,
  Rowan-Robinson, Rubino-Martin, Rusholme, Said, Salvatelli, Salvati, Sandri,
  Santos, Savelainen, Savini, Scott, Seiffert, Serra, Shellard, Spencer,
  Spinelli, Stolyarov, Stompor, Sudiwala, Sunyaev, Sutton, Suur-Uski, Sygnet,
  Tauber, Terenzi, Toffolatti, Tomasi, Tristram, Trombetti, Tucci, Tuovinen,
  Turler, Umana, Valenziano, Valiviita, Van~Tent, Vielva, Villa, Wade, Wandelt,
  Wehus, White, White, Wilkinson, Yvon, Zacchei, \&
  Zonca}]{Collaboration:2016bk}
Planck~Collaboration, P., Ade, P. A.~R., Aghanim, N., {et~al.} 2016, Astronomy
  {\&} Astrophysics, 594, A13

\bibitem[{{Pol} {et~al.}(2019){Pol}, {McLaughlin}, \& {Lorimer}}]{Pol2019A}
{Pol}, N., {McLaughlin}, M., \& {Lorimer}, D.~R. 2019, \apj, 870, 71

\bibitem[{Punturo {et~al.}(2010)Punturo, Abernathy, Acernese, Allen, Andersson,
  Arun, Barone, Barr, Barsuglia, Beker, Beveridge, Birindelli, Bose, Bosi,
  Braccini, Bradaschia, Bulik, Calloni, Cella, Chassande-Mottin, Chelkowski,
  Chincarini, Clark, Coccia, Colacino, Colas, Cumming, Cunningham, Cuoco,
  Danilishin, Danzmann, De~Luca, De~Salvo, Dent, De~Rosa, Di~Fiore,
  Di~Virgilio, Doets, Fafone, Falferi, Flaminio, Franc, Frasconi, Freise,
  Fulda, Gair, Gemme, Gennai, Giazotto, Glampedakis, Granata, Grote, Guidi,
  Hammond, Hannam, Harms, Heinert, Hendry, Heng, Hennes, Hild, Hough, Husa,
  Huttner, Jones, Khalili, Kokeyama, Kokkotas, Krishnan, Lorenzini, L{\"u}ck,
  Majorana, Mandel, Mandic, Martin, Michel, Minenkov, Morgado, Mosca, Mours,
  M{\"u}ller-Ebhardt, Murray, Nawrodt, Nelson, O{\textquoteright}Shaughnessy,
  Ott, Palomba, Paoli, Parguez, Pasqualetti, Passaquieti, Passuello, Pinard,
  Poggiani, Popolizio, Prato, Puppo, Rabeling, Rapagnani, Read, Regimbau,
  Rehbein, Reid, Rezzolla, Ricci, Richard, Rocchi, Rowan, R{\"u}diger,
  Sassolas, Sathyaprakash, Schnabel, Schwarz, Seidel, Sintes, Somiya, Speirits,
  Strain, Strigin, Sutton, Tarabrin, Th{\"u}ring, van~den Brand, van Leewen,
  van Veggel, Van Den~Broeck, Vecchio, Veitch, Vetrano, Vicer{\'e}, Vyatchanin,
  Willke, Woan, Wolfango, \& Yamamoto}]{Punturo:2010jf}
Punturo, M., Abernathy, M., Acernese, F., {et~al.} 2010, Classical and Quantum
  Gravity, 27, 194002

\bibitem[{Reitze {et~al.}(2019)}]{Reitze:2019dyk}
Reitze, D., {et~al.} 2019, arXiv:1903.04615

\bibitem[{{Safarzadeh} \& {Berger}(2019)}]{SB19}
{Safarzadeh}, M., \& {Berger}, E. 2019, arXiv e-prints, arXiv:1904.08436

\bibitem[{Safarzadeh {et~al.}(2019)Safarzadeh, Ramirez-Ruiz, Andrews, Macias,
  Fragos, \& Scannapieco}]{Safarzadeh:2019dd}
Safarzadeh, M., Ramirez-Ruiz, E., Andrews, J.~J., {et~al.} 2019, The
  Astrophysical Journal, 872, 105

\bibitem[{Safarzadeh {et~al.}(2018)Safarzadeh, Sarmento, \&
  Scannapieco}]{Safarzadeh:2018ub}
Safarzadeh, M., Sarmento, R., \& Scannapieco, E. 2018, eprint arXiv:1812.02779,
  1812.02779

\bibitem[{{Sathyaprakash} {et~al.}(2012)}]{sathya:ETcqg}
{Sathyaprakash}, B., {et~al.} 2012, Classical and Quantum Gravity, 29, 124013

\bibitem[{Sathyaprakash {et~al.}(2009)Sathyaprakash, Schutz, \&
  Broeck}]{Sathyaprakash:2009jc}
Sathyaprakash, B.~S., Schutz, B., \& Broeck, C. V.~D. 2009, Classical and
  Quantum Gravity, 215006

\bibitem[{{Sathyaprakash} \& {Schutz}(2009)}]{sathya:gwreview}
{Sathyaprakash}, B.~S., \& {Schutz}, B.~F. 2009, Living Reviews in Relativity,
  12, 2

\bibitem[{{Schutz}(2011)}]{schutz:antenna}
{Schutz}, B.~F. 2011, Classical and Quantum Gravity, 28, 125023

\bibitem[{Schutz(2011)}]{Schutz:2011fn}
Schutz, B.~F. 2011, Classical and Quantum Gravity, 125023

\bibitem[{Shen {et~al.}(2015)Shen, Cooke, Ramirez-Ruiz, Madau, Mayer, \&
  Guedes}]{Shen:2015gc}
Shen, S., Cooke, R.~J., Ramirez-Ruiz, E., {et~al.} 2015, The Astrophysical
  Journal, 807, 115

\bibitem[{Simonetti {et~al.}(2019)Simonetti, Matteucci, Greggio, \&
  Cescutti}]{Simonetti:2019uq}
Simonetti, P., Matteucci, F., Greggio, L., \& Cescutti, G. 2019, eprint
  arXiv:1901.02732, 1901.02732

\bibitem[{Taylor \& Gair(2012)}]{Taylor:2012jo}
Taylor, S.~R., \& Gair, J.~R. 2012, Physical Review D, 86, 2

\bibitem[{Taylor {et~al.}(2011)Taylor, Gair, \& Mandel}]{Taylor:2011ex}
Taylor, S.~R., Gair, J.~R., \& Mandel, I. 2011, Physical Review D, 688

\bibitem[{Usman {et~al.}(2018)Usman, Mills, \& Fairhurst}]{Usman:2018tv}
Usman, S.~A., Mills, J.~C., \& Fairhurst, S. 2018, 1809.10727

\bibitem[{van~de Voort {et~al.}(2015)van~de Voort, Quataert, Hopkins, Kere{\v
  s}, \& Faucher-Gigu{\`e}re}]{vandeVoort:2015jw}
van~de Voort, F., Quataert, E., Hopkins, P.~F., Kere{\v s}, D., \&
  Faucher-Gigu{\`e}re, C.-A. 2015, Monthly Notices of the Royal Astronomical
  Society, 447, 140

\bibitem[{Vigna-G{\'o}mez {et~al.}(2018)Vigna-G{\'o}mez, Neijssel, Stevenson,
  Barrett, Belczynski, Justham, de~Mink, M{\"u}ller, Podsiadlowski, Renzo,
  Sz{\'e}csi, \& Mandel}]{VignaGomez:2018th}
Vigna-G{\'o}mez, A., Neijssel, C.~J., Stevenson, S., {et~al.} 2018, eprint
  arXiv:1805.07974, 1805.07974

\bibitem[{{Vitale} \& {Farr}(2018)}]{VitaleFarr18}
{Vitale}, S., \& {Farr}, W.~M. 2018, arXiv e-prints, arXiv:1808.00901

\bibitem[{{Vitale} \& {Whittle}(2018)}]{2018PhRvD..98b4029V}
{Vitale}, S., \& {Whittle}, C. 2018, \prd, 98, 024029

\bibitem[{Zheng \& Ramirez-Ruiz(2007)}]{Zheng:2007hl}
Zheng, Z., \& Ramirez-Ruiz, E. 2007, The Astrophysical Journal, 665, 1220

\end{thebibliography}
